\documentclass[12pt,thmsa]{article}

\input{tcilatex}

\begin{document}

\begin{center}
{\Large Topological-Torsion and Topological-Spin}

{\Large \ as coherent structures in plasmas}

\bigskip

R. M. Kiehn

\textit{Physics Department, University of Houston}

rkiehn2352@aol.com \ \ 

http://www.cartan.pair.com

\bigskip
\end{center}

\textbf{Abstract} \ The PDE's of classical electromagnetism can be generated
from two exterior differential systems that distinguish topologically the
field intensities and potentials, $F-dA=0,$\ from the field excitations and
the charge current densities, $J-dG=0$. \ The existence of potentials, $A$,
leads to the independent 3-forms of T-Torsion = $A\symbol{94}F$ and T-Spin = 
$A\symbol{94}G$. \ The exterior derivatives (divergences) of these 3-forms
produce anomalies that define the two classic Poincare invariants. The
closed integrals of these forms, when deformation invariants of frozen-in
fields, define topological coherent structures in the plasma. \ \vspace{1pt}%
Solutions when T-Torsion (T-Spin) is zero define transverse magnetic
(electric) modes on topological grounds. \ \vspace{1pt}When the divergence
of  T-Torsion is not zero there exists a classical mechanism for charge
acceleration along the magnetic field lines producing symplectic plasma
currents; large temperature gradients along the $\mathbf{B}$ field lines can
act as a source of stellar plasma jets in neutron stars. \ In such
circumstances, the Torsion vector is uniquely defined by conformal
invariance of the Action potentials. \ Plasma currents in the direction of
the Torsion vector leave both $A\symbol{94}F$ and $A\symbol{94}G$
conformally invariant, hence these fields are frozen-in even though the
processes are thermodynamically irreversible. The decaying coherent and
deformable topological structures associated with such frozen-in fields are
persistent and observable artifacts, similar to wakes, that can appear in
any plasma domain, such as that which surrounds stars.

\section{\protect\vspace{1pt}Introduction}

In the language of exterior differential systems [1]\ it becomes evident
that classical electromagnetism is equivalent to a set of topological
constraints on a variety of independent variables. \ Certain integral
properties of an electromagnetic system are deformation invariants with
respect to all continuous evolutionary processes that can be described by a
singly parameterized vector field. \ These deformation invariants lead to
the fundamental topological conservation laws described in the physical
literature as the conservation of charge and the conservation of flux. \
\medskip Recall the definitions:

\begin{quote}
A continuous process is defined as a map from an initial state of topology $%
T_{initial}$ into a final state of perhaps different topology $T_{final}$
such that the limit points of the initial state are permuted among the limit
points of the final state. [2]\ \medskip

A deformation invariant is defined as an integral over a closed manifold, $%
\int_{z}\omega $ such that the Lie derivative of the closed integral with
respect to a singly parameterized vector field, $\beta V^{k},\ $\ vanishes,
for any choice of parametrization, $\beta $. \ \medskip
\end{quote}

\[
L_{(\beta V^{k})}\int_{z}\omega =0\;\;\;\;any\;\beta 
\]
The idea of a deformation invariant comes from the Cartan concept of a tube
of trajectories as applied to Hamiltonian mechanics. \ Consider the flow
lines tangent to the trajectories generated by $V^{k},$ and a closed
integration chain that connects points on a tube of different trajectories.
\ Under certain conditions (when the virtual work vanishes)\ the integral of
the exterior 1-form of Action, $A=pdq-H(p,q,t)dt$\ evaluated along the
closed integration chain yields a value which is the same no matter how the
integration chain is deformed, as long as it resides on the same tube of
trajectories. \ As the points on the trajectories have relative
displacements determined by a factor $\beta (p,q,t)$ then the closed chain
connecting points can be deformed by choosing a different function $\beta .$
\ Cartan used this idea for demonstrating that the tube of trajectories is
uniquely defined on a contact manifold by a Hamiltonian flow that conserves
energy. \lbrack 3\rbrack\ \ He thereby defined conservative Hamiltonian
processes in a topological manner by requiring that processes be the subsets
of singly parameterized vector fields that leave the closed integral of the
1-form of Action a deformation invariant. \ 

\qquad However, for physical systems that can be defined by a 1-form of
Action, $A$, the derived 2-form $F=dA$ \ is a deformation invariant with
respect to \textit{all} continuous processes that can be defined by a singly
parameterized vector field. \ This concept is at the basis of the Helmholtz
theorems in hydrodynamics, and the conservation of flux in classical
electromagnetism. \ \ Herein, this topological constraint will be called the
postulate of potentials. \ When written as the equation, $F-dA=0,$ the
postulate of potentials is to be recognized as an exterior differential
system constraining the topology of the independent variables. \ From Stokes
theorem, the (2 dimensional) domain of finite support for $F$ can not\ in
general be compact without boundary, unless the Euler characteristic
vanishes. \ There are two exceptional cases for two dimensional domains, the
(flat or twisted) torus and the Klein-Bottle, but these situations require
the additional topological constraint that $F\symbol{94}F=0.$\ The fields in
these exceptional cases must reside on these exceptional compact surfaces,
which form topological coherent structures in the electromagnetic field. \
For an electromagnetic action, the exceptional compact cases can only exist
if $\mathbf{E\circ B}=0.$\ The resulting statement is that there do not
exist compact domains of support without boundary when $\mathbf{E\circ B}%
\neq 0,\,$a statement that will be of interest to thermodynamics of
irreversible systems, and of plasma jets.

The definition of an electromagnetic system of charges and currents will
require a second topological constraint imposed upon the domain of
independent variables. This second postulate will be called the postulate of
conserved currents. The electromagnetic domain not only supports the 1-form $%
A$, but also supports an N-1 form density, $J,$which is exact. The
equivalent differential system, $J-dG=0,\,$requires that the (N-1
dimensional) domain of support for $J$ cannot be compact without boundary. \
However, the closed integrals of $J$ are deformation invariants for \textit{%
any} continuous evolutionary process that can be defined in terms of a
singly parameterized vector field.

In section 2, the classical Maxwell system will be displayed in terms of the
vector formalism of Sommerfeld and Stratton. \ The key feature is to note
that the fields of intensities ($\mathbf{E}$ and $\mathbf{B}$) are
considered as separate and distinct from the fields of excitation ($\mathbf{D%
}$ and $\mathbf{H}$), a historical distinction (championed by Sommerfeld)
that is often masked in modern exposes of electromagnetic theory. \ 

In section 3, it will be demonstrated explicitly that the classic formalism
of electromagnetism in section 2 is a consequence of a system of two
fundamental topological constraints

\[
F-dA=0,\,\,\,\,\,\,\,\,\,\ J-dG=0. 
\]
defined on a domain of four independent variables. The theory requires the
existence of four fundamental exterior differential forms, $\{A,F,G,J\},$
which can be used to construct the complete Pffaf sequence [4]\ of forms by
the processes of exterior differentiation and exterior multiplication. \ On
a domain of four independent variables, the complete Pfaff sequence contains
three 3-forms: the classic 3-form of charge current density, $J,$and the
(apparently novel to many researchers) \ 3-forms of Topological Spin Current
density, $A\symbol{94}G,$[5]\ and Topological Torsion-Helicity, $A\symbol{94}%
F$\ \ [6]. \ To shorten notation, the terms T-Torsion and T-Spin will be
used.

As the charge current 3-form, $J,$ is a deformation invariant by
construction, it is of interest to determine topological refinements or
constraints for which the 3-forms of \ T-Spin and T-Torsion will define
physical topological conservation laws in the form of deformation
invariants. \ The additional constraints are equivalent to the topological
statement that the closure (exterior derivative) of each of the three forms
is empty (zero). \ It will be demonstrated in section 4 that these closure
conditions define the two classic Poincare invariants (4-forms) as
deformation invariants, and when each of these invariants vanish the
corresponding 3-form generates a topological quantity (T-Spin or T-Torsion
respectively) which is also a deformation invariant. \ The possible values
of the topological quantities, as deRham period integrals [7], form rational
ratios.

The concepts of \ T-Spin Current and the T-Torsion vector have been utilized
hardly at all in applications of classical electromagnetic theory. \ Just as
the vanishing of the 3-form of charge current, $J=0,$ defines the
topological domain called the vacuum, the vanishing of the two other 3-forms
will refine the fundamental topology of the Maxwell system.\ \ \ Such
constraints permit a definition of transversality to be made on topological
(rather than geometrical) grounds. If both $A\symbol{94}G\,\ $and $A\symbol{%
94}F$ vanish, the vacuum state supports topologically transverse modes only
(TTEM). \ Examples lead to the conjecture that TTEM modes do not transmit
power, a conjecture that has been verified when the concept of geometric
transversality (TEM) and topological transversality (TTEM) coincide. \ \ A
topologically transverse magnetic (TTM) mode corresponds to the topological
constraint that $A\symbol{94}F=0.$\ A topologically transverse electric mode
(TTE) corresponds to the topological constraint that $A\symbol{94}G=0.$\ \
Examples, both novel and well-known, of vacuum solutions to the
electromagnetic system which satisfy (and which do not satisfy) these
topological constraints are given in the appendix. \ These ideas should be
of interest to those working in the field of Fiber Optics. \ Recall that
classic solutions which are geometrically and topologically transverse (TEM$%
\equiv $TTEM) do not transmit power [8]. \ \ However, in the appendix an
example vacuum wave solution is given which is geometrically transverse (the
fields are orthogonal to the field momentum and the wave vector), and yet
the geometrically transverse wave transmits power at a constant rate: \ the\
example wave is not topologically transverse as $A\symbol{94}F\neq 0.$\ 

In section 4, an additional topological constraint will be used to define
the plasma process as a restriction on all processes which can be described
in terms of a singly parameterized vector field. \ \ The plasma process
(which is to be distinguished from a Hamiltonian process) will be restricted
to those vector fields which leave the closed integrals of $G$ a deformation
invariant. \ (Compare to the Cartan definition that a Hamiltonian process is
a restriction on arbitrary processes such that the closed integrals of $A$
are deformation invariants with respect to Hamiltonian processes). \ \ A\
plasma process need not conserve energy. \ A \textit{perfect} plasma process
is a plasma process which is also a Hamiltonian process. \ Again, the three
forms, $J,\;$$A\symbol{94}G$ and $A\symbol{94}F$ are of particular
interested for their tangent manifolds define ''lines'' in the 4-dimensional
variety of space and time. \ Relative to plasma processes, the topological
evolution associated with such lines, and their entanglements, is of utility
in understanding solar corona and plasma instability. [9]\medskip

\section{The Domain of Classical Electromagnetism}

\subsection{\protect\vspace{1pt}The classical Maxwell-Faraday and the
Maxwell-Ampere equations.}

Using the notation and the language of Sommerfeld and Stratton [10], the
classic definition of an electromagnetic system is a domain of space-time $%
\{x,y,z,t\}$ which supports both the Maxwell-Faraday equations,

\begin{equation}
\vspace{1pt}curl\,\,\mathbf{E}+\partial \mathbf{B}/\partial
t=0,\,\,\,\,\,\,\,\,\,\,div\,\,\mathbf{B}=0,
\end{equation}
and the Maxwell-Ampere equations,

\begin{equation}
curl\,\,\mathbf{H}-\partial \mathbf{D}/\partial t=\mathbf{J}%
,\,\,\,\,\,\,\,\,\,\,div\,\,\mathbf{D}=\rho .
\end{equation}

\subsection{The conservation of charge current}

\ In every case, the charge current density for the Maxwell system satisfies
the conservation law,

\begin{equation}
div\,\mathbf{J}+\partial \rho /\partial t=0.
\end{equation}
The charge-current densities are subsumed to be zero $\lbrack \mathbf{J}$, $%
\rho \rbrack =0$ for the vacuum state. \ For the Lorentz vacuum state, the
field excitations, $\mathbf{D}$ and $\mathbf{H}$, are linearly connected to
the field intensities, $\mathbf{E}$ and $\mathbf{B}$, \ by means of the
Lorentz (homogeneous and isotropic) constitutive relations:

\begin{equation}
\mathbf{D}=\varepsilon \mathbf{E}\,,\,\,\,\,\,\mathbf{B}=\mu \mathbf{H.}
\end{equation}

The two vacuum constraints imply that the solutions to the homogeneous
Maxwell equations also satisfy the vector wave equation, typically of the
form

\begin{equation}
grad\,\,div\,\,\mathbf{B}-curl\,curl\mathbf{\,B}-\varepsilon \mu \partial
^{2}\mathbf{B}/\partial t^{2}=0.
\end{equation}
The constant wave phase velocity,$\,\,\,v_{p},$is taken to be

\begin{equation}
v_{p}^{2}=1/\varepsilon \mu \equiv c^{2}
\end{equation}
Similar results can be obtained for the solid state where the constitutive
constraints can be more complex [11], and for the plasma state where the
charge-current densities are not zero.

\subsection{The existence of potentials}

It is further subsumed that the classic Maxwell electromagnetic system is
constrained by the \ statement that the field intensities are deducible from
a system of twice differentiable potentials, $[\mathbf{A},\phi ]$:

\begin{equation}
\mathbf{B}=curl\,\mathbf{\,A,\,\,\,\,\,\,\,\,E}=-grad\,\,\phi -\partial 
\mathbf{A}/\partial t.
\end{equation}
This constraint topologically implies that domains that support non-zero
values for the covariant field intensities, $\mathbf{E}$ and $\mathbf{B,}$%
can \textit{not} be compact domains without a boundary. \ It is this
constraint that distinguishes classical electromagnetism from Yang Mills
theories. \ Two other classical 3-vector fields are of interest, the
Poynting vector $\mathbf{E\times H}$ representing the flux of
electromagnetic radiative energy, and the field momentum flux, $\mathbf{%
D\times B}.$

\section{\protect\vspace{1pt}The Fundamental Exterior Differential Systems.}

\vspace{1pt}The formulation of Maxwell theory in section 2 is relative to a
choice of independent variables $\{x,y,z,t\}$using classical vector analysis
developed in euclidean 3-space.\ The topological features of the formalism
are not immediately evident. \ However, electromagnetism has a formulation
in terms of Cartan's exterior differential forms [12]. \ Exterior
differential forms do not depend upon a choice of coordinates, do not depend
upon the a choice of metric, and are independent of the constraints imposed
by gauge groups and connections. \ In such a formulation the equations of an
electromagnetic system become recognized as consequences of topological
constraints on a domain of independent variables.

The use of differential forms should not be viewed as just another formalism
of fancy. \ The technique goes beyond the methods of tensor calculus, and
admits the study of topological evolution. \ Recall that if an exterior
differential system is valid on a final variety of independent variables
\{x,y,z,t\}, then it is also true on any initial variety of independent
variables that can be mapped onto \{x,y,z,t\}. The map need only be
differentiable, such that the Jacobian matrix elements are well defined 
\textit{functions}. \ The Jacobian matrix does not have to have an inverse,
so that the exterior differential system is not restricted to the
equivalence class of diffeomorphisms. \ The field intensities on the initial
variety are functionally well defined by the pullback mechanism, which
involves algebraic composition with components of the Jacobian matrix
transpose, and the process of functional substitution. \ This independence
from a choice of independent variables (or coordinates) for Maxwell's
equations was first reported by Van Dantzig [13]. \ It follows that the\
Maxwell differential system is well defined in a covariant manner for both
Galilean transformations as well as Lorentz transformations, or any other
diffeomorphism. \ (The singular solution sets to the equations do not enjoy
this universal property). In addition, it should be noted that the ideas of
the exterior differential system imply that the closure equations of the
Maxwell-Faraday type form a nested set, with exactly the same format,
independent of the choice of the \textit{number} of independent variables. \
In addition, every physical system (such as fluid) that supports a 1-form of
Action, has its version of the Maxwell-Faraday induction equations.\medskip

\subsection{The Maxwell-Faraday exterior differential system.}

\vspace{1pt}The Maxwell-Faraday equations are a consequence of the exterior
differential system

\begin{equation}
F-dA=0,
\end{equation}
where $A$ is a 1-form of Action, with twice differentiable coefficients
(potentials proportional to momenta) which induce a 2-form, $F,$of
electromagnetic intensities ($\mathbf{E}$ and $\mathbf{B\,,\,\,\,}$related
to forces and objects of intensities$)$. \ The exterior \ differential
system is a topological constraint that in effect defines field intensities
in terms of the potentials. On a four dimensional space-time of independent
variables, $(x,y,z,t)$ the 1-form of Action (representing the postulate of
potentials) can be written in the form

\begin{equation}
A=\Sigma _{k=1}^{3}A_{k}(x,y,z,t)dx^{k}-\phi (x,y,z,t)dt=\mathbf{A\circ }d%
\mathbf{r-}\phi dt.
\end{equation}

Subject to the constraint of the exterior differential system, the 2-form of
field intensities, $F,$becomes:

\begin{eqnarray}
F &=&dA=\{\partial A_{k}/\partial x^{j}-\partial A_{j}/\partial x^{k}\}dx^{j}%
\symbol{94}dx^{k} \\
&=&F_{jk}dx^{j}\symbol{94}dx^{k}=\mathbf{B}_{z}dx\symbol{94}dy...\mathbf{E}%
_{x}dx\symbol{94}dt...
\end{eqnarray}
where in usual engineering notation,

\begin{equation}
\mathbf{E}=-\partial \mathbf{A}/\partial t-grad\phi \mathbf{%
,\,\,\,\;\;\;\;\;B=}curl\,\,\mathbf{A}\equiv \partial A_{k}/\partial
x^{j}-\partial A_{j}/\partial x^{k}.
\end{equation}
The closure of the exterior differential system, $dF=0,\,$

\begin{equation}
dF=ddA=\{curl\;\mathbf{E}+\partial \mathbf{B}/\partial t\}_{x}dy\symbol{94}dz%
\symbol{94}dt-..+..-div\,\mathbf{B}dx\symbol{94}dy\symbol{94}dz\}\Rightarrow
0,
\end{equation}
generates the Maxwell-Faraday partial differential equations.:

\begin{equation}
\{curl\;\mathbf{E}+\partial \mathbf{B}/\partial t=0,\,\,\,\,\,\,\,\,\,\,div\,%
\mathbf{B}=0\}.
\end{equation}
The component functions ($\mathbf{E}$ and $\mathbf{B})\,$of the 2-form, $F,$%
transform as covariant tensor of rank 2. \ \ The topological constraint that 
$F$\ is exact, implies that the domain of support for the field intensities
cannot be compact without boundary, unless the Euler characteristic
vanishes. These facts distinguish classical electromagnetism from Yang-Mills
field theories (where the domain of support for $F$ is presumed to be
compact without boundary). \ Moreover, the fact that $F$ is subsumed to be
exact and C1 differentiable excludes the concept of magnetic monopoles from
classical electromagnetic theory on topological grounds. \ The closed
integral of the 2-form $F$ over any closed 2-manifold is a deformation
(topological)\ invariant of any evolutionary process that can be described
by a singly parameterized vector field, for

\begin{eqnarray}
\vspace{1pt}L_{\mathbf{V}}(\int_{z2}F) &=&\int_{z2}\{i(V)dF+d(i(V)F)\}= \\
\int_{z2}\{0+d(i(V)F)\} &=&\int_{z2}d(i(V)F)=0
\end{eqnarray}
The integral is then a deformation invariant, for the result is valid even
if the 4-vector field is distorted by an arbitrary function, $%
f\{x,y,z,t\},\, $such that $\mathbf{V}$$\Rightarrow f(x,y,z,t)\mathbf{V}.$ \
The notation $\int_{z2}$implies that the 2D integration chain is closed. \
It can be a cycle or \vspace{1pt}a boundary.

\subsection{The Maxwell Ampere exterior differential system}

\qquad The Maxwell Ampere equations are a consequence of second exterior
differential system,

\begin{equation}
J-dG=0,
\end{equation}
where $G$ is an N-2 form \textit{density} of field excitations ($\mathbf{D}$
and $\mathbf{H\,,\,\,\,}$related to sources or objects of quantity$)$, and $%
J $ is the N-1 form of charge-current densities. \ The partial differential
equations equivalent to the exterior differential system are precisely the
Maxwell-Ampere equations. \ This second postulate, on a four dimensional
domain of independent variables, assumes the existence of a N-2 form density
given by the expression,

\begin{equation}
G=G^{34}(x,y,z,t)dx\symbol{94}dy...+G^{12}(x,y,z,t)dz\symbol{94}dt...=%
\mathbf{D}^{z}dx\symbol{94}dy...\mathbf{H}^{z}dz\symbol{94}dt...
\end{equation}
Exterior differentiation produces an N-1 form,

\begin{equation}
J=\mathbf{J}^{z}(x,y,z,t)dx\symbol{94}dy\symbol{94}dt...-\rho (x,y,z,t)dx%
\symbol{94}dy\symbol{94}dz.\ 
\end{equation}

Matching the coefficients of the exterior expression $dG=J$\ leads to the
Maxwell-Ampere equations,

\begin{equation}
curl\,\mathbf{H}-\partial \mathbf{D/}\partial t=\mathbf{J\,}%
\,\,\,\,\,\,\,and\,\,\,\,\,\mathbf{\,\,}div\,\mathbf{D}=\rho .
\end{equation}
The fact that $J$ is exact leads to the charge conservation law, $dJ=ddG=0,$%
or

\begin{equation}
\partial \mathbf{J}^{x}/\partial x+\partial \mathbf{J}^{y}/\partial
y+\partial \mathbf{J}^{z}/\partial z+\partial \rho /\partial t=0.
\end{equation}

The exterior differential system is a topological constraint for by Stokes
theorem the support for $G$ can be compact without boundary only if the
domain is without charge-currents. \ The closure of the exterior
differential system, $dJ=0,$ generates the charge-current conservation law.
\ The integral of $J$ over a closed 3 dimensional domain is a relative
integral invariant (a deformation invariant) of any process that can be
described in terms of a singly parametrized vector field. \ The formal
statement is given by Cartan's magic formula [14], which describes
continuous topological evolution in terms of the action of the Lie
derivative, with respect to a vector field, acting on the exterior
differential 3-form, $J:$

\begin{equation}
\vspace{1pt}L_{\mathbf{V}}(\int_{z3}J)=\int_{z3}\{i(V)dJ+d(i(V)J)\}=%
\int_{z3}\{0+d(i(V)J)\}=0.
\end{equation}
The Lie derivative of the closed integral is equal to zero for any 4-vector
field $V,$when $dJ=0.$\ The integral is then a deformation invariant, for
the result is valid even if the 4-vector field is distorted by an arbitrary
function, $f\{x,y,z,t\},\,$such that $\mathbf{V}$$\Rightarrow f(x,y,z,t)%
\mathbf{V}. $\ 

\subsection{\protect\vspace{1pt}\ The T-Torsion and T-Spin 3-forms}

\qquad As mentioned above, the method of exterior differential forms goes
beyond the domain of classical tensor analysis, for it admits of maps from
initial to final state that are without inverse. \ (Tensor analysis and
coordinate transformations require that the Jacobian map from initial to
final state has an inverse - the method of exterior differential forms does
not.) Hence the theory of electromagnetism expressed in the language of
exterior differential forms admits of topological evolution, at least with
respect to continuous processes without Jacobian inverse. \ With respect to
such non-invertible maps, both tensor fields and differential forms are not
functionally well defined in a predictive sense \lbrack 15\rbrack . \ Given
the functional forms of a tensor field on an initial state, it is impossible
to predict uniquely the functional form of the tensor field on the final
state unless the map between initial and final state is invertible. \
However differential forms are functionally well defined in a retrodictive
sense, by means of the pullback. \ Covariant anti-symmetric tensor fields
pull back retrodictively with respect to the transpose of the Jacobian
matrix (of functions) and functional substitution, and contravariant tensor
densities pullback retrodictively with respect to the adjoint of the
Jacobian matrix, and functional substitution. \ The transpose and the
adjoint of the Jacobian exist, even if the Jacobian inverse does not.

The exterior differential forms that make up the electromagnetic system
consist of the primitive 1-form, $A$, and the primitive N-2 form density, $%
G, $their exterior derivatives, and their algebraic intersections defined by
all possible exterior products. The complete Maxwell system of exterior
differential forms (the Pfaff sequence for the Maxwell system) is given by
the set:

\begin{equation}
\{A,F=dA,G,\,\,J=dG,\,A\symbol{94}F,\,A\symbol{94}G,\,A\symbol{94}J,\,F%
\symbol{94}F,\,G\symbol{94}G\}.
\end{equation}
These forms and their unions may be used to form a topological base on the
domain of independent variables. \ \ The Cartan topology constructed on this
system of forms has the useful feature that the exterior derivative may be
interpreted as a limit point, or closure, operator in the sense of
Kuratowski [16]. The exterior differential systems that define the
Maxwell-Ampere and the Maxwell-Faraday equations above are essentially
topological constraints of closure. Note that the complete Maxwell system of
differential forms (which assumes the existence of $A$) also generates two
other exterior differential systems.\qquad

\begin{equation}
d(A\symbol{94}G)-(F\symbol{94}G-A\symbol{94}J)=0,
\end{equation}
and

\begin{equation}
d(A\symbol{94}F)-F\symbol{94}F=0.
\end{equation}
The two objects, $A\symbol{94}G$ and $A\symbol{94}F$ are three forms, not
usually found in discussions of classical electromagnetism. \ The closed
components of the first 3-form (density) were called T-Spin [17] and the
second 3-form, were called T-Torsion (or helicity) [18]. \ By direct
evaluation of the exterior product, and on a domain of 4 independent
variables, each 3-form will have 4 components that can be symbolized by the
4-vector arrays

\begin{equation}
Spin-Current:\mathbf{S}_{4}=[\mathbf{A\times H}+\mathbf{D}\phi ,\mathbf{%
A\circ D}]\equiv \lbrack \mathbf{S,}\sigma \mathbf{]},
\end{equation}
and

\begin{equation}
Torsion-vector:\mathbf{T}_{4}=[\mathbf{E\times A}+\mathbf{B}\phi ,\mathbf{%
A\circ B}]\equiv \lbrack \mathbf{T,}h\mathbf{]},
\end{equation}
which are to be compared with the charge current 4-vector density:

\begin{equation}
Ch\arg e-Current:\mathbf{J}_{4}=[\mathbf{J},\rho ],
\end{equation}
The 3-forms then can be defined by the equivalent contraction processes

\begin{eqnarray}
&&Topological\,\,\,Spin\,\,\,3-form\doteq \,\,A\symbol{94}G \\
&=&i(\mathbf{S}_{4})dx\symbol{94}dy\symbol{94}dz\symbol{94}dt=\mathbf{S}%
^{x}dy\symbol{94}dz\symbol{94}dt.....-\sigma dx\symbol{94}dy\symbol{94}dz
\end{eqnarray}
and

\begin{eqnarray}
&&Topological\,\,Torsion-helicity\,\,\,\,3-form\doteq A\symbol{94}F \\
&=&i(\mathbf{T}_{4})dx\symbol{94}dy\symbol{94}dz\symbol{94}dt=\,\,\mathbf{T}%
^{x}dy\symbol{94}dz\symbol{94}dt.....-hdx\symbol{94}dy\symbol{94}dz.
\end{eqnarray}

The vanishing of the first 3-form is a topological constraint on the domain
that defines topologically transverse electric (TTE) waves: the vector
potential, $\mathbf{A}$, is orthogonal to $\mathbf{D,}$ in the sense that $%
\mathbf{A\circ D}=0.$\ The vanishing of the second 3-form is a topological
constraint on the domain that defines topologically transverse magnetic
(TTM)\ waves: \ the vector potential, $\mathbf{A}$, is orthogonal to $%
\mathbf{B,\,} $\ in the sense that $\mathbf{A\circ B}=0.$\ \ When both
3-forms vanish, the topological constraint on the domain defines
topologically transverse (TTEM) waves. \ For classic real fields this double
constraint would require that vector potential, $\mathbf{A,}$ is collinear
with the field momentum, $\mathbf{D\times B,}$ and in the direction of the
wave vector, $\mathbf{k}$.$\mathbf{\,\,\,} $

The geometric notion of distinct transversality modes of electromagnetic
waves is a well known concept experimentally, but the association of
transversality to topological issues is novel herein. For certain examples
that appear in the appendix, it is apparent that the concept of geometric
and topological transversality are the same. \ \ In the classic case, often
considered in fiber optic theory, it is known that the TEM modes do transmit
power. However, in the appendix, a vacuum wave solution is given which
satisfies the geometric concept of transversality ( it is both a TM and a TE
solution) but the mode radiates for it is not both a TTM and a TTE solution.
\ The conjecture obtained from examples is that a TTEM solution does not
radiate.

Note that if the 2-form $F\,$\ was not exact, such topological concepts of
transversality would be without meaning, for the 3-forms of T-Spin and
T-Torsion depend upon the existence of the 1-form of Action. \ The torsion
vector $\mathbf{T}_{4}$ and theT-Spin vector $\mathbf{S}_{4}$ are associated
vectors to the 1-form of Action in the sense that

\begin{equation}
i(\mathbf{T}_{4})A=0\,\,\,\ and\,\,\,\,\,i(\mathbf{S}_{4})A=0
\end{equation}

\subsection{The Poincare Invariants}

The exterior derivatives of the 3-forms of \ T-Spin and T-Torsion produce
two 4-forms, $F\symbol{94}G-A\symbol{94}J$ and $F\symbol{94}F,$\ whose
integrals over closed 4 dimensional domains are deformation invariants for
the Maxwell system. \ \ These topological objects\ are related to the
conformal invariants of a Lorentz system as discovered by Poincare and
Bateman. \ In the format of independent variables $\{x,y,z,t\},$ the
exterior derivative corresponds to the 4-divergence of the 4-component
T-Spin and T-Torsion vectors, $\mathbf{S}_{4}\,\ $and $\mathbf{T}_{4}.$\ The
functions so created define the Poincare conformal invariants of the Maxwell
system:

\begin{eqnarray}
Poincare\,\,1 &=&d(A\symbol{94}G)=F\symbol{94}G-A\symbol{94}J \\
&=&\{div_{3}(\mathbf{A\times H}+\mathbf{D}\phi )+\partial (\mathbf{A\circ D)}%
/\partial t\}dx\symbol{94}dy\symbol{94}dz\symbol{94}dt \\
&=&\{(\mathbf{B\circ H-D\circ E)-(A\circ J}-\rho \phi )\}dx\symbol{94}dy%
\symbol{94}dz\symbol{94}dt
\end{eqnarray}

\begin{eqnarray}
Poincare\,\,2 &=&d(A\symbol{94}F)=F\symbol{94}F \\
&=&\{div_{3}(\mathbf{E\times A}+\mathbf{B}\phi )+\partial (\mathbf{A\circ B)}%
/\partial t\}dx\symbol{94}dy\symbol{94}dz\symbol{94}dt \\
&=&\{-2\mathbf{E\circ B}\}dx\symbol{94}dy\symbol{94}dz\symbol{94}dt
\end{eqnarray}

For the vacuum state, with $J=0,$ zero values of the Poincare invariants
require that the magnetic energy density is equal to the electric energy
density $(1/2\mathbf{B\circ H}=1/2\mathbf{D\circ E)}$, and, respectively,
that the electric field is orthogonal to the magnetic field $(\mathbf{E\circ
B}=0\mathbf{).}$\ Note that these constraints often are used as elementary
textbook definitions of \ what is meant by electromagnetic waves. \ When
either Poincare invariant vanishes, the corresponding closed 3-dimensional
integral becomes a topological quantity in the sense of a deRham period
integral. \ \ For example, when the first Poincare invariant vanishes, the
closed integral of the 3-form ofT-Spin becomes a deformation invariant with
quantized values:

\begin{eqnarray}
Define\,Topo\log ical\,\,Spin &=&\int_{z3}A\symbol{94}G\text{ \ \ \ \ \ \ }
\\
\text{\ }Let\text{ }d(A\symbol{94}G) &=&0,\,then \\
\vspace{1pt}L_{\mathbf{V}}\,(Spin) &=&\int_{z3}\{i(V)d(A\symbol{94}%
G)+d(i(V)(A\symbol{94}G)\} \\
&=&\int_{z3}\{0+d(i(V)(A\symbol{94}G)\}=0.
\end{eqnarray}

Similarly, when the second Poincare invariant vanishes, the closed integral
of the 3-form of \ T-Torsion-Helicity becomes a deformation invariant with
quantized values:

\begin{eqnarray}
\vspace{1pt}Define\ Topo\log ical\,\,Torsion &=&\int_{z3}A\symbol{94}F\text{%
\thinspace \thinspace \thinspace \thinspace \thinspace \thinspace \thinspace
\thinspace \thinspace \thinspace \thinspace\ } \\
Let\text{\thinspace \thinspace \thinspace }d(A\symbol{94}F)
&=&0,\,\,\,\,\,\,then \\
L_{\mathbf{V}}(Torsion\text{-}Helicity) &=&\int_{z3}\{i(V)d(A\symbol{94}%
F)+d(i(V)(A\symbol{94}F)\} \\
&=&\int_{z3}\{0+d(i(V)(A\symbol{94}F)\}=0.
\end{eqnarray}
It is important to realize that these topological conservation laws are
valid in a plasma as well as in the vacuum, subject to the conditions of
zero values for the Poincare invariants. \ On the other hand, topological
transitions require that the Poincare invariants are not zero.

\section{Deformation Invariants and the Plasma State.}

\subsection{\protect\vspace{1pt}Special evolutionary processes. \ The plasma
process}

As described in a previous section, the fundamental equation of topological
evolution is given by Cartan's magic formula, which acts as a propagator on
the forms that make up the exterior differential system. \ As stated in the
first paragraph, an evolutionary\ process is defined herein as a map that
can be described by a singly parameterized vector field. \ If the Action of
the Lie derivative on the complete system of Maxwell exterior differential
forms vanishes for a particular choice of process, then that process leaves
the entire Maxwell system absolutely invariant. \ As a topology can be
constructed in terms of an exterior differential system, \ and if a special
process leaves that system of forms invariant, then the topology induced by
the system of forms is invariant; \ the process must be a homeomorphism.

However, for a given Maxwell system, it is more likely that only some of the
exterior differential forms that make up the Maxwell system are invariant
relative to an arbitrary process; others are not. Of particular interest are
those forms which are relative integral invariants of continuous
deformations. \ The closed integral of the form is not only invariant with
respect to a process represented by particular vector field, but also with
respect to longitudinal deformations of that process obtained by multiplying
the particular vector field by an arbitrary function. \ For vector fields
which are singly parameterized, this concept of longitudinal deformation is
equivalent to a reparameterization of the vector field. \ 

\qquad The development that follows is guided by Cartan's pioneering work,
in which he examined those specialized processes for mechanical systems that
leave the 1-form of Action, $A,$ a deformation invariant. \ Cartan proved
that such processes always have a Hamiltonian representation. \ An
electromagnetic system has not only the primitive 1-form, $A,$ but also the
N-2 form, $G,$ which can undergo evolutionary processes. \ For
electromagnetic systems, a particular interesting choice of specialized
processes are those that leave the N-2 form, $G,$ of field excitations a
deformation (relative) integral invariant. \ The equations that must be
satisfied are of the form

\begin{eqnarray}
L_{\beta \mathbf{V}}(\int_{z2}G) &=&\int_{z2}i(\beta V)dG=\int_{z2}i(\beta
V)J \\
&=&\int_{z2}\beta \{(\mathbf{J}-\rho \mathbf{V)}^{x}dy\symbol{94}dz-...+(%
\mathbf{J}\times \mathbf{V})^{x}dx\symbol{94}dt...\Rightarrow 0
\end{eqnarray}
It follows that deformation invariance of the N-2 form $G$\ requires that
the admissable evolutionary processes be restricted to those that satisfy
the definitions of the classical plasma:

\begin{equation}
\mathbf{J}=\rho \mathbf{V.}
\end{equation}
\ 

(This constraint is used to define the ''Plasma state''in this article). \
As the closed integrals of $G$ are by Gauss law, the counters of net charge
within the closed domain, the classical plasma equation is to be recognized
as the statement that in the closed domain the net number of charges is a
deformation invariant. \ That is, charges can be produced only in equal and
opposite pairs by a ''plasma process''. \ A plasma process does not involve
net charge production. \ 

This invariance principle is to be compared to the Helmholtz theorem which
checks on the validity of the deformation integral invariance of the 2-form $%
F.$

\begin{equation}
L_{\beta \mathbf{V}}(\int_{z2}F)=\int_{z2}i(\beta V)dF=0
\end{equation}
The closed integral of Helmholtz is an intrinsic topological (deformation)
invariant of an electromagnetic system, for the 2-form $F$ is exact by
construction (the postulate of potentials).\ The Helmholtz integral is a
deformation invariant for all evolutionary processes that can be described
by a singly parameterized vector field. \ (This statement is not true for
Yang Mills fields). \ \ Hence in a plasma, for which the evolutionary
processes are constrained such that $\mathbf{J}=\rho \mathbf{V}$, both the
closed integrals of$\ F$ and $G$ are deformation invariants. \ In the sense,
the plasma is a topological refinement of the complete Maxwell system.

In the subsections that follow, various topological categories of plasma
processes will be examined. \ The ideal and semi-ideal plasma processes will
obey the plasma master equation, and the non-ideal plasma processes will
not. \ The electromagnetic flux is a local (absolute) invariant of all
semi-ideal plasma processes. \ This statement is similar to the
classification of hydrodynamic flows. \ Ideal and semi-ideal hydrodynamic
flows satisfy the Helmholtz theorem, and the local conservation of vorticity.

\subsection{\protect\medskip The ideal plasma process = a Hamiltonian
process.}

\qquad Next consider the evolutionary properties of the 1-form of Action in
the plasma state by evaluating the possible deformation invariance of the
1-form of Action, $A,$ with respect to motions that preserve the plasma
state:

\begin{equation}
L_{\rho \mathbf{V}}(\int_{z1}A)=\int_{z1}i(\rho
V)dA=\int_{z1}W=\int_{z1}\{(\rho \mathbf{E}+\mathbf{J\times B})_{k}dx^{k}+(%
\mathbf{J\circ E})dt\}\Rightarrow 0.
\end{equation}
The 1-form $\ W$ is the 1-form of virtual work defined in terms of the
Lorentz force. \ The resulting equation demonstrates that the concept of a
Lorentz force, $\rho \mathbf{E}+\mathbf{J\times B,}$has a topological
foundation. \ It is apparent that if the Lorentz force vanishes, 
\begin{equation}
\{\rho \mathbf{E}+\mathbf{J\times B})\Rightarrow 0,
\end{equation}
and the Plasma current density is NOT ohmic, 
\begin{equation}
(\mathbf{J\circ E})=\rho (\mathbf{V\circ E})\Rightarrow 0,
\end{equation}
$\,$then the closed integral of the Action 1-form is also a deformation
topological invariant of the Plasma process. \ Such a set of constraints, 
\begin{equation}
W=i(\rho V)dA=0,
\end{equation}
topologically defines the ''ideal''\ plasma state as a plasma process for
which the 1-form of virtual Work vanishes. \ By Cartan's theorem, the 1-form
of Action then has a unique Hamiltonian representation and the ideal plasma
process is uniquely defined as a Hamiltonian process ( the Pfaff dimension
of the 1-form, $A$ must processes be 3 or less for uniqueness). \ The ideal
plasma is thereby a restriction of arbitrary processes to that unique
process that leaves invariant both the closed integrals of flux and the
closed integrals of charge. \ \ Ideal plasmas are electromagnetic systems
for which the admissable processes are the intersection of a plasma process
and a unique Hamiltonian process. The ideal plasma can not exist on a domain
of a 4 dimensional variety where the second Poincare invariant is not zero.$%
\ $\ 

\subsection{The Bernoulli-Casimir plasma process is a semi-ideal plasma
process.}

The topological constraint that the 1-form of virtual work vanishes is
sufficient but not necessary for a \ plasma process to preserve the closed
integrals of the Action 1-form. \ Evolutionary invariance of the closed
integral of Action does not require that the plasma process be unique. \ The
1-form of virtual Work, $W,\,$need not be zero, but only closed: \ $%
dW\Rightarrow 0.$\ By analogy to hydrodynamics, if the virtual Work 1-form
is exact,

\begin{equation}
\vspace{1pt}W=d\Theta
\end{equation}
then the Lorentz force is represented by a spatial gradient, $\rho \mathbf{E}%
+\mathbf{J\times B}=\nabla \Theta ,\,$\ and the Power $-\mathbf{J\circ E}%
=\partial \Theta /\partial t.$ \ The function $\Theta (x,y,z,t)$ is a
Bernoulli-Casimir function, and acts as the generator of a symplectic
Hamiltonian flow. \ The (non-unique) Bernoulli-Casimir function is an
evolutionary invariant for each process path, but is not necessarily a
constant over the domain:

\begin{equation}
L_{\rho \mathbf{V}}(\Theta )=i(\rho V)d\Theta =i(\rho V)i(\rho V)A=0.
\end{equation}
The Bernoulli-Casimir function is not the same as the Hamiltonian energy
function, but is more closely related to the thermodynamic concept of
enthalpy. \ The Bernoulli-Casimir function can be used to generate a
''Hamiltonian process'', but the process is not uniquely defined.

For such symplectic plasma processes, the gradient of the Bernoulli-Casimir
function is transverse to the $\mathbf{B}$ field only when the second
Poincare invariant vanishes.

\begin{equation}
\medskip \rho \,\mathbf{E\circ B}=\nabla \Theta \circ \mathbf{B.}
\end{equation}
Similar expression were studied in conjunction with topological conservation
in MHD by Hornig and Schindler [20].

\begin{equation}
\medskip \rho \,\mathbf{E\circ V}=\nabla \Theta \circ \mathbf{V.}
\end{equation}

If the Ohmic assumption is made for the plasma process, $\mathbf{J}=\rho 
\mathbf{V}=\sigma (\mathbf{E}+\mathbf{V\times B})\,$, then the symplectic
condition leads to a thermopower format of the type

\begin{equation}
\mathbf{J=(}1/\rho \sigma )grad(kT)
\end{equation}
when it is subsumed that the Bernoulli-Casimir function is related to
temperature. It would appear that for plasma motion along the $\mathbf{B}$
field lines, there can exist a dynamo action to produce an $\mathbf{E}$
field collinear with the magnetic field. \ 

\qquad It is suggested that the large temperature gradient that exists in a
plasma envelope about a rotating star (with a $\mathbf{B}$ field like a
neutron star) can induce a current flow and an $\mathbf{E}$ acceleration
field along the polar \ magnetic field lines. \ Like in a Bernoulli process
in a fluid, the mechanical energy is not conserved, but the enthalpy (the
Bernoulli-Casimir) is a invariant along any trajectory, and that invariant
can be different from trajectory to neighboring trajectory.

\subsection{The Stokes plasma process is a semi-ideal process that obeys the
Master equation.}

The constraint that the virtual work 1-form, $W$, generated by a plasma
process, $W=i(\rho V)dA$, be closed, does not require that it be exact. \
The constraint of closure yields two vector conditions:

\begin{equation}
dW=0\Rightarrow curl(\rho \mathbf{E}+\mathbf{J\times B)}=0\,\,\,\,\,\,\,\,\,%
\,\,\,and\,\,\,\,\,\,\,\,\,\,\,\nabla (\mathbf{J\circ E)=}\partial (\rho 
\mathbf{E}+\mathbf{J\times B})/\partial t.
\end{equation}
The first vector condition implies that

\begin{equation}
\nabla \rho \times (\mathbf{E+V\times B)}+\rho \,curl(\mathbf{E)}+curl(%
\mathbf{V\times B)}=0.
\end{equation}
By using the Maxwell-Faraday equation, this topological constraint becomes
the plasma master equation:

\begin{equation}
-\partial \mathbf{B}/\partial t+curl(\mathbf{V\times B)=-}\nabla \ln \rho
\times (\mathbf{E+V\times B).}
\end{equation}
All of these ideal and semi ideal plasma processes enjoy the property that
the electromagnetic flux is conserved locally. \ That is

\begin{equation}
L_{\rho \mathbf{V}}(dA)=L_{\rho \mathbf{V}}F=d(i(\rho V)F)=0.
\end{equation}
\medskip

\subsection{\ Frozen-in lines.}

It is of some interest to examine the evolution of the differential forms
that make up an electromagnetic system relative to Plasma processes. \ The
method is to construct the Lie derivative with respect a plasma process, $%
\mathbf{J}=\rho \mathbf{V,}$of all forms that make up the electromagnetic
Pfaff sequence.

For an arbitrary vector field Z whose tangents define a line in space time,
the N-1 form

\begin{equation}
W=i(\gamma Z)dx\symbol{94}dy\symbol{94}dz\symbol{94}dt
\end{equation}
can be tested for evolutionary invariance relative to any other vector $V.$\
Suppose the effect of the evolutionary process is conformal:

\begin{equation}
L_{(V)}W=i(V)d\,W+d(i(V)W)=\Gamma (x,y,z,t)W
\end{equation}
This statement implies that the points that make up the tangent line of the
vector field W remain on the tangent line. \ The points may be permuted but
they do not leave the line. \ Such is the concept of a frozen in field. \
The points that make up a line evolve into points of the same line. \ The
evolution need not be uniformly continuous, especially where the points are
folded. \ Yet even in such cases the points of a line are still points of
the line, even though rearranged in order. \ If for a given $V$ the
evolution of the lines of $W$ is conformal, then there exists a
parametrization of $V$\ such that the evolution is uniform and invariant. \
A parametrization function $\beta (x,y,z,t)\,$\ can be found such that

\begin{equation}
L_{(\beta V)}W=\beta L_{(V)}W+L_{(V)}\beta \symbol{94}W=(\beta \cdot \Gamma
\,+i(V)d\beta )W\Rightarrow 0.
\end{equation}

\vspace{1pt}For the electromagnetic system there are three N-1 forms, which
may or may not be frozen into the evolutionary process. \ Consider the
3-form of current.

\begin{equation}
L_{(V)}J=i(V)d\,J+d(i(V)J)
\end{equation}
As $dJ=0,$

\begin{equation}
L_{(V)}J=d\{i(V)i(J)dx\symbol{94}dy\symbol{94}dz\symbol{94}dt\}
\end{equation}
It follows that if $i(V)i(J)dx\symbol{94}dy\symbol{94}dz\symbol{94}dt\}=0,$
the field lines of $J$ are frozen-in (with $\Gamma =0)$. \ So the plasma
evolutionary process evolutionary, with $\mathbf{J}=\rho \mathbf{V}$, is an
example of a process that ''freezes-in'' the lines of current. \ \ However,
there are many other evolutionary processes for which the $J$ lines are
frozen in. \ 

The formulas created by 4.16 are valid on any set of independent variables,
but expressions on 4 dimensions of space time for ''frozen-in'' lines are
not quite the same as those that appear in the engineering literature based
on euclidean 3-space [21]. \ Either the time-like component of the 4-vector $%
W$ must vanish, or the process $V$ must be explicitly time-independent for
the general formulas to be in precise agreement with the engineering
expressions. [22]

\qquad It is important to note that in space time the\ ''frozen-in'' lines
must be related to 3-forms, and not to the two form components $\mathbf{E}$
and $\mathbf{B}.$\ These latter objects can produce ''frozen-in'' lines only
on the exceptional 2-surfaces, the torus and the Klein bottle, (and then
only when $\mathbf{E}$$\circ $$\mathbf{B=}0)\mathbf{.}$\ The 3-form of \
T-Torsion has spatial components that are dominated by the $\mathbf{B}$
field (in the limit $\mathbf{E}\rightarrow 0)$, such that ''frozen-in''
lines of T-Torsion might have the appearance of ''frozen-in'' lines of $%
\mathbf{B}.$\ The 3-form of T-Spin has lines that can be dominated by $%
\mathbf{A\times H.}$ \ However the explicit formulas for the 3-forms of
T-Torsion and T-Spin are not dependent upon a choice of constitutive
relations that act as geometrical constraints on the 2-forms of $F$ and $G.$%
\ See below.

\subsection{Evolution of the lines of T-Torsion with respect to plasma
currents.}

\vspace{1pt}Consider the evolution of the lines of T-Torsion

\begin{eqnarray}
L_{(\rho V)}A\symbol{94}F &=&i(\rho V)d(A\symbol{94}F)+d(i(\rho V)A\symbol{94%
}F) \\
&=&i(\rho V)d(A\symbol{94}F)+d\{(i(\rho V)A)\symbol{94}F-A\symbol{94}i(\rho
V)F\}
\end{eqnarray}

First consider those systems where the second Poincare invariant \ vanishes, 
$F\symbol{94}F=0.$\ The lines in space time which are tangent to the 3-form $%
A\symbol{94}F$\ \ then have zero divergence. \ The lines can only start and
stop on boundary points, or they are closed on themselves. \ The T-Torsion
lines can be either parallel to the plasma current or they can be orthogonal
to the plasma current. \ As the electromagnetic current is exact, any three
dimensional domain of support for a finite plasma current cannot be compact
without a boundary. \ \ If the lines of plasma current start and stop on
boundary points, then the lines of T-Torsion can form closed loops that link
the current lines. \ \ It is the concept of linkages that is of interest to
the theory of magnetic knots.

\qquad Consider that plasma process such that the evolution is in the
direction of the T-Torsion lines. \ As in this situation,

\begin{eqnarray}
(i(J)A\symbol{94}F) &=&(i(\rho V)A\symbol{94}F)\Rightarrow (i(\gamma \mathbf{%
T}_{4})A\symbol{94}F) \\
&=&\gamma (i(\mathbf{T}_{4})(i(\mathbf{T}_{4})dx\symbol{94}dy\symbol{94}dz%
\symbol{94}dt=0,
\end{eqnarray}
the 3-form of T-Torsion is a local invariant whenever the second Poincare
invariant vanishes; $\mathbf{E\circ B}\Rightarrow 0$. \ In other words, $F%
\symbol{94}F\neq 0$ is a local necessary condition for topological change. \
It is also a remarkable fact that any evolution in the direction of the
Torsion vector leaves the Action 1-form conformally invariant, in the sense
that:

\begin{equation}
L_{(\gamma \mathbf{T}_{4})}A=i(\gamma \mathbf{T}_{4})dA+di(\gamma \mathbf{T}%
_{4})A=\gamma (\mathbf{E\circ B})A+0.
\end{equation}
The torsion vector on a domain of 4 variables is transverse to the 1-form of
Action, as $A\symbol{94}(A\symbol{94}F)=0.$\ Evolution in the direction of
the Torsion vector in not Hamiltonian, unless the second Poincare invariant
vanishes. \ In section 6 below this idea will be related to thermodynamic
irreversibility.

\subsection{Evolution of the lines of T-Spin Current with respect to plasma
currents.}

Consider the evolution of the lines ofT-Spin current

\begin{eqnarray}
L_{(\rho V)}A\symbol{94}G &=&i(\rho V)d(A\symbol{94}G)+d(i(\rho V)A\symbol{94%
}G) \\
&=&i(\rho V)d(A\symbol{94}G)+d\{(i(\rho V)A)\symbol{94}G-A\symbol{94}i(\rho
V)G\}
\end{eqnarray}

First consider those systems where the first Poincare invariant \ vanishes, $%
F\symbol{94}G-A\symbol{94}J=0.$\ The lines in space time which are tangent
to the 3-form $A\symbol{94}G$then have zero divergence. \ The lines can only
start and stop on boundary points, or they are closed on themselves. \ The
T-Spin lines are either parallel to the plasma current or they are
orthogonal to the plasma current. \ As the electromagnetic current is exact,
any three dimensional domain of support for a finite plasma current cannot
be compact without a boundary. \ \ If the lines of plasma current do not
stop or start on boundary points (current loops), then the T-Spin lines
which terminate on boundary points can be linked by the current loops.

The concept of the T-Spin vector depends on the existence of $G$, but not on
the concept of $J=dG$. \ That is, the T-Spin vector can be associated with
separated domains of charges, which can be compact domains without boundary
that are compliments of the domain of finite charge current densities, which
are domains that can not be compact without boundary.

\section{Thermodynamics}

\subsection{\protect\vspace{1pt}Topological Thermodynamics and
Irreversibility}

\vspace{1pt}The basic tool for studying topological evolution is Cartan's
magic formula, in which it is presumed that a physical (hydrodynamic) system
can be described adequately by a 1-form of Action, $A$, and that a physical
process can be represented by a contravariant vector field, $\mathbf{V}$,
which can be used to represent a dynamical system or a flow:\ 
\begin{eqnarray}
L_{(\mathbf{V})}\tint A &=&\tint L_{(\mathbf{V})}A=\tint \{i(\mathbf{V}%
)dA+d(i(\mathbf{V})A)\}  \label{6} \\
&=&\tint \{W+d(U)\}=\tint Q\smallskip .
\end{eqnarray}

The basic idea behind this formalism (which is at the foundation of the
Cartan-Hilbert variational principle) is that postulate of potentials is
valid: $F-dA=0$. \ The base manifold will be the 4-dimensional variety $%
\{x,y,z,t\}$ of engineering practice, but no metrical features are presumed
a priori. If relative to the process, $V$, the RHS of equation \ref{(6)} is
zero, $\tint Q\smallskip .\Rightarrow 0$, then $\tint A$ is said to be an
integral invariant of the evolution generated by $\mathbf{V.}$\ \ \ In
thermodynamics such processes are said to be adiabatic.

From the point of view of differential topology, the key idea is that the
Pfaff dimension, or class [23], of the 1-form of Action specifies
topological properties of the system. Given the Action 1-form, $A$, the
Pfaff sequence, $\{A,dA,A\symbol{94}dA,dA\symbol{94}dA,...\}$ will terminate
at an integer number of terms $\leq $the number of dimensions of the domain
of definition. On a 2n+2=4 dimensional domain, the top Pfaffian, $dA\symbol{%
94}dA$, will define a volume element with a density function whose singular
zero set (if it exists) reduces the symplectic domain to a contact manifold
of dimension 2n+1=3. This (defect) contact manifold supports a unique
extremal field that leaves the Action integral ''stationary'', and leads to
the Hamiltonian conservative representation for the Euler flow in
hydrodynamics. \ The irreversible regime will be on an irreducible
symplectic manifold of Pfaff dimension 4, where $dA\symbol{94}dA\neq 0.$\
Topological defects (or coherent structures) appear as singularities of
lesser Pfaff (topological) dimension, $dA\symbol{94}dA=0.$\ \ 

\qquad Classical hydrodynamic processes can be represented by certain nested
categories of vector fields, $\mathbf{V}$. \ Recall that in order to be
Extremal, the process, $\mathbf{V}$, must satisfy the equation\medskip

\begin{equation}
Extremal\,--(unique\,\,Hamiltonian):\,\,\,\,\,\ \ \
\,\,\,\,\,\,\,\,\,\,\,\,\,i(\mathbf{V})dA=0;\medskip  \label{5.2a}
\end{equation}
in order to be Hamiltonian the process must satisfy the equation\medskip

\begin{equation}
Bernouilli--Casimir--Hamiltonian:\,\,\,\,\,\ \ \,\,i(\mathbf{V})dA=d\Theta
;\medskip  \label{5.2b}
\end{equation}
in order to be Symplectic, the process must satisfy the equation\medskip

\begin{equation}
Helmholtz--Symplectic:\,\,\,\,\,\,\,\,\,\,\,\,\,\,\,\,\,\,\,\,\,\,\,\,\,\,\,%
\,\,\,\,\,\,\,\,\,\,\,\,\,\,\,\,\,\,\,\,\,\,\,\,\,\,\,\,di(\mathbf{V}%
)dA=0.\medskip  \label{5.2c}
\end{equation}

Extremal processes cannot exist on the non-singular symplectic domain,
because a non-degenerate anti-symmetric matrix (the coefficients of the
2-form $dA)$ does not have null eigenvectors on space of even dimensions . \
Although unique extremal stationary states do not exist on the domain of
Pffaf dimension 4, there can exist evolutionary invariant Bernoulli-Casimir
functions, $\Theta ,$ that generate non-extremal, ''stationary''states. Such
Bernoulli processes can correspond to energy dissipative symplectic
processes, but they, as well as all symplectic processes, are reversible in
the thermodynamic sense described below. The mechanical energy need not be
constant, but the Bernoulli-Casimir function(s), $\Theta ,$are evolutionary
invariant(s), and may be used to describe non-unique stationary state(s). \ 

The equations, above, that define several familiar categories of processes,
are in effect constraints on the topological evolution of any physical
system represented by an Action 1-form, $A.$\ The Pfaff dimension of the
1-form of virtual work, $W=i(\mathbf{V})dA$ is 1 or less for the three
categories. \ The extremal constraint of equation \ref{5.2a} can be used to
generate the Euler equations of hydrodynamics for a incompressible fluid.
The Bernoulli-Casimir constraint of equation \ref{5.2b} can be used to
generate the equations for a barotropic compressible fluid. \ The Helmholtz
constraint of equation \ref{5.2c} can be used to generate the equations for
a Stokes flow. \ All such processes are thermodynamically reversible. \ \
None of these constraints above will generate the Navier-Stokes equations,
which require that the topological dimension of the 1-form of virtual work
must be greater than 2. \ 

A crucial idea is the recognition that irreversible processes must on
domains of Pfaff dimension \ \ which support T-Torsion, $A\symbol{94}dA\neq
0,$ with its attendant properties of non-uniqueness, envelopes, regressions,
and projectivized tangent bundles. Such domains are of Pfaff dimension 3 or
greater. \ Moreover, as described below, it would appear that thermodynamic
irreversibility must support a non-zero Topological Parity 4-form, \ $dA%
\symbol{94}dA\neq 0.$ \ Such domains are of Pfaff dimension 4 or greater. \ 

Although there does not exist a unique gauge independent stationary state on
the symplectic manifold of Pfaff dimension 4, remarkably there does exist a
unique vector field on the symplectic domain, with components that are
generated by the 3-form $A\symbol{94}dA$. This unique (to within a factor)
vector field is defined as the T-Torsion vector, $\mathbf{T}_{4}$, and
satisfies (on the 2n+2=4 dimensional manifold) the equation,\medskip

\begin{equation}
i(\mathbf{T}_{4})dx\symbol{94}dy\symbol{94}dz\symbol{94}dt=A\symbol{94}%
dA\medskip
\end{equation}
This (four component) vector field, $\mathbf{T}_{4}$, has a non-zero
divergence almost everywhere, for if the divergence is zero, then the 4-form 
$dA\symbol{94}dA$ vanishes, and the domain is no longer a symplectic
manifold! The T-Torsion vector, $\mathbf{T}_{4}$, can be used to generate a
dynamical system that will decay to the stationary states $(div_{4}(\mathbf{T%
}_{4}\mathbf{)}\Rightarrow 0)$ starting from arbitrary initial conditions.
These processes are irreversible in the thermodynamic sense. \ It is
remarkable that this unique evolutionary vector field, $\mathbf{T}_{4}$, is
completely determined (to within a factor) by the physical system itself;
e.g., the components of the 1-form, $A$, determine the components of the
T-Torsion vector.

To understand what is meant by thermodynamic irreversibility, realize that
Cartan's magic formula of topological evolution is equivalent to the first
law of thermodynamics.\medskip

\begin{equation}
L_{(\mathbf{v})}A=i(\mathbf{V})dA+d(i(\mathbf{V})A)=W+dU=Q.\medskip
\end{equation}
$A$ is the ''Action'' 1-form that describes the hydrodynamic system. $%
\mathbf{V} $ is the vector field that defines the evolutionary process. $W$
is the 1-form of (virtual) work. $Q$ is the 1-form of heat. From classical
thermodynamics, a process is irreversible when the heat 1-form $Q$ does not
admit an integrating factor. From the Frobenius theorem, the lack of an
integrating factor implies that $Q\symbol{94}dQ\neq 0.$ \ Hence a simple
test may be made for any process, $\mathbf{V}$, relative to a physical
system described by an Action 1-form, $A$:\medskip

\begin{equation}
If\,\,\,\,\,L_{(\mathbf{v})}A\symbol{94}L_{(\mathbf{v})}dA\neq
0\,\,then\,\,the\,\,process\,\,is\,\,irreversible.\medskip
\end{equation}

This topological definition implies that the three categories (above) of
symplectic, Hamiltonian or extremal processes, $\subset \mathbf{S,}$ are
reversible$\,\,(as\ L_{(\mathbf{S})}dA=dQ=0).\,$\ However, for evolution in
the direction of the T-Torsion vector, $\mathbf{T}_{4}$, direct computation
demonstrates that the fundamental equations lead to a conformal evolutionary
process, \ a process which is thermodynamically irreversible:\medskip 

\begin{equation}
\,\,L_{(\mathbf{T}_{4})}A=\sigma A\,\,\,\,\,\,\,\,\,and\,\,\,\,\,\,\,i(%
\mathbf{T}_{4})A=0,\medskip
\end{equation}
such that\medskip

\begin{equation}
L_{(\mathbf{T}_{4})}A\symbol{94}L_{(\mathbf{T}_{4})}dA=Q\symbol{94}dQ=\sigma
^{2}A\symbol{94}dA\neq 0.\medskip
\end{equation}
It is remarkable for the irreversible case that the Lie derivative with
respect to $T$ acting on $A$ is comparable to the covariant derivative. \
However the Lie derivative with respect to $T$ acting on $dA$ is not
equivalent to a covariant derivative.

\subsection{\protect\medskip Applications to Electromagnetism and Plasmas}

All of the development of previous sections will carry over to the
electromagnetic system, which also subsumes the postulate of potentials. \
The T-Torsion 3-form, $A\symbol{94}dA$, induces the T-Torsion vector,

\begin{equation}
\mathbf{T}_{4}=\{\mathbf{(E\times A+B}\phi );\mathbf{A\circ B}\}\equiv \{%
\mathbf{S},h\}.\medskip
\end{equation}

\vspace{1pt}If $div_{4}\mathbf{T}=-2\,\,\mathbf{E\circ B}\neq 0,$ the
electromagnetic 1-form, $A,$ defines a domain of Pfaff dimension 4. \ Such
domains cannot support topologically transverse magnetic waves $(as\,\,\,A%
\symbol{94}F\neq 0)$. \ Evolutionary processes (including plasma currents)
that are proportional to the T-Torsion vector are thermodynamically
irreversible, if $\,\,\sigma =\mathbf{E\circ B}\neq 0$. \ However, the
conformal properties of evolution in the direction of the T-Torsion vector
lead to extraordinary properties when the plasma current is in the direction
of the T-Torsion vector. \ From the thermodynamic arguments presented above,
based on the postulate of potentials for an arbitrary system, but using the
notation of an electromagnetic system, \ it follows that

\begin{equation}
L_{(\mathbf{T}_{4})}A=\sigma A=-(\mathbf{E}\circ \mathbf{B})A
\end{equation}
and

\begin{equation}
L_{(\mathbf{T}_{4})}(A\symbol{94}F)=2\sigma A=-2(\mathbf{E}\circ \mathbf{B}%
)\;A\symbol{94}F.
\end{equation}
It follows that motion along the direction of the torsion vector freezes-in
the lines of the torsion vector in space time, but the process is
irreversible unless the second Poincare invariant is zero. \ The time
evolution of the deformable coherent structure is recognizable even though
it thermodynamically decays!

Recall that the definition of a plasma current, $J,$ is equivalent to an
evolutionary process such that

\begin{equation}
\text{Definition\thinspace \thinspace of\thinspace a \thinspace plasma
\thinspace Current}\,\,\,J\,:\,\,\,\,\,\,\,\,\,\,\,L_{(J)}G=0.
\end{equation}
Consider a plasma current which is also in the direction of the Torsion
vector. \ Then

\begin{eqnarray}
L_{(J)}A\symbol{94}G &=&(L_{(J)}A)\symbol{94}G+A\symbol{94}L_{(J)}G \\
&=&(L_{(\gamma \mathbf{T}_{4})}A)\symbol{94}G+A\symbol{94}L_{(\gamma \mathbf{%
T}_{4})}G=\gamma \cdot (\mathbf{E\circ B})\,\,\,A\symbol{94}G+0
\end{eqnarray}
For plasma motions in the direction of the (possibly dissipative) torsion
vector, both the ''lines''of theT-Spin vector are ''frozen in'' and the
lines of the Torsion vector are ''frozen in''. \ Such ''frozen in''objects
can be used to give a topological definition of deformable coherent
structures in a plasma. \ Moreover, as the evolutionary process causes the
frozen in structures to deform and decay, it is conceivable that evolution
could proceed to form stationary ( but not stagnant) states (where $\mathbf{E%
}\circ \mathbf{B}\Rightarrow 0),$ such that the frozen in field line
structures become local deformation invariants, or topological defects. \ 

\vspace{1pt}In conclusion, electromagnetic coherent structures are
evolutionary deformable (and perhaps decaying) domains of Pfaff dimension 4,
which form stationary states of topological defects (including the null
state) \ in regions of Pfaff dimension 3, where $\,\mathbf{E\circ B}=0$.\ \
(Note that all semi-ideal plasma current processes are reversible in a
thermodynamic sense.)

\section{Electromagnetic Waves in the Vacuum with T-Spin and T-Torsion}

As the T-Spin 4-vector and the T-Torsion 4-vector formalism may be
unfamiliar to many readers, it is useful to compare four classes of unusual
vacuum wave solutions with the usual waveguide solutions. \ The ''unusual
waves'' have their vector potential, $\mathbf{A},$ orthogonal to the wave
vector, $\mathbf{k}$, describing the direction of the wave front. \ In each
unusual example, the current density is in the direction of the vector
potential and therefore also orthogonal to the wave vector. The usual wave
solutions have their vector potential parallel to the wave vector. \ The
four unusual cases belong to equivalence classes defined by the constraints

\begin{eqnarray*}
(A\symbol{94}F &=&0,A\symbol{94}G\neq 0) \\
(A\symbol{94}F &\neq &0,A\symbol{94}G=0) \\
(A\symbol{94}F &=&0,A\symbol{94}G=0) \\
(A\symbol{94}F &\neq &0,A\symbol{94}G\neq 0).
\end{eqnarray*}
In each case, each component of the potentials satisfies the wave equation
subject to the phase velocity relation, ($\omega /k)^{2}-1/(\xi \mu )=0.$ \
The current density, $J,$ is proportional to the vector potential, $A$, (in
a fashion reminiscent to the London conjecture) multiplied by the same phase
velocity relation. \ The examples do not generate any charge current
distributions when the phase velocity equation is satisfied (the phase
velocity equals the speed of light as determined by the constitutive
equations). \ 

\vspace{1pt}In each example given below, the 1-form of Action is specified
and the field intensities are computed. \ Then the T-Spin Current and the
T-Torsion vector are evaluated. \ The functions have been chosen to satisfy
the Lorentz vacuum conditions of zero charge current densities, subject to a
phase velocity ''dispersion'' relation. \ The phase function is defined by
the formula $\Theta =(\pm kz\mp wt)$ representing outbound eaves.\ The
Poynting vector is computed, and the Poincare invariants are evaluated. \ 

\medskip Examples of the four classes of these simple (but unusual)\ wave
types correspond to:

\subsection{Example 1. Real Linear Polarization:}

Consider the Potentials

\begin{equation}
A=[\cos (kz-\omega t),\cos (kz-\omega t),0,0]\,\,\,
\end{equation}

and their induced fields:

\[
\mathbf{E}=\lbrack -\sin (kz-\omega t),-\sin (kz-\omega t),0\rbrack \omega 
\]

\[
\mathbf{B}=\lbrack +\sin (kz-\omega t),-\sin (kz-\omega t),0\rbrack k 
\]

\[
\mathbf{J}_{4}=\lbrack \cos (kz-\omega t),\cos (kz-\omega t),0,0\rbrack
(k^{2}-\varepsilon \mu \omega ^{2})/\mu 
\]

\[
\mathbf{S}_{4}=\lbrack 0,0,-k/\mu ,-\varepsilon \omega \rbrack \,2\cos
(kz-\omega t)\sin (kz-\omega t). 
\]

\[
\mathbf{T}_{4}=\lbrack 0,0,0,0\rbrack . 
\]

\[
\mathbf{E\times H\,}\,=\lbrack 0,0,1\rbrack (\,\omega k/\mu )(2\cos
(kz-\omega t)^{2}-1) 
\]

\begin{eqnarray*}
(\mathbf{B\circ H-D\circ E)} &=&-2\{\cos (kz-\omega t)^{2}-\sin (kz-\omega
t)^{2}\}(k^{2}-\varepsilon \mu \omega ^{2})/\mu \mathbf{\,\,} \\
\mathbf{\,\,\,\,\,\,\,\,\,\,\,\,\,\,\,\,\,\,\,\,\,\,(E\circ B)} &=&0
\end{eqnarray*}

This class of potentials generates a set of complex field intensities and
excitations, and a current density proportional to the vector potential. \
If the dispersion relation $(k^{2}-\varepsilon \mu \omega ^{2})=0$ is
satisfied, then the solutions are acceptable vacuum solutions, with a
vanishing charge current density. \ The T-Torsion vector vanishes
identically, independent from the dispersion condition, but theT-Spin vector
does not. \ The first Poincare invariant vanishes subject to the constraint
of the dispersion relation. \ The second Poincare invariant vanishes
identically. \ The solution corresponds to a linear state of polarization at
45$^{\circ }$ with respect to the x-axis, with the electric and the magnetic
fields in phase. \ There is a non-zero Poynting vector along the z axis.,
which is orthogonal to the vector potential. \ Note that the radiated power
has a time average which is zero. If the charge current density is not zero
(due to a fluctuation in the dispersion relation) the charge current vector
is orthogonal to theT-Spin current vector.

\subsection{Example 2. \ Real Circular Polarization:}

Consider the Potentials

\begin{equation}
A=[\cos (kz-\omega t),\sin (kz-\omega t),0,0]\,\,
\end{equation}
and their induced fields:

\[
\mathbf{E}=\lbrack -\sin (kz-\omega t),+\cos (kz-\omega t),0\rbrack \omega 
\]

\[
\mathbf{B}=\lbrack -\cos (kz-\omega t),-\sin (kz-\omega t),0\rbrack k 
\]

\[
\mathbf{J}_{4}=\lbrack \cos (kz-\omega t),\sin (kz-\omega t),0,0\rbrack
(k^{2}-\varepsilon \mu \omega ^{2})/\mu 
\]

\[
\mathbf{S}_{4}=\lbrack 0,0,0,0\rbrack . 
\]

\[
\mathbf{T}_{4}=\lbrack 0,0,-\omega ,-k\rbrack . 
\]

\[
\,\,\mathbf{E\times H\,}=\lbrack 0,0,1\rbrack \,\omega \,k/\mu 
\]

\[
(\mathbf{B\circ H-D\circ E)}=0\mathbf{\,\,\,\,\,\,\,\,\,\,\,\,\,\,\,\,\,\,\,%
\,\,\,\,\,(E\circ B)}=0 
\]

This class of potentials generates a set of complex field intensities and
excitations, and a current density proportional to the vector potential. \
If the dispersion relation $(k^{2}-\varepsilon \mu \omega ^{2})=0$ is
satisfied, then the solutions are acceptable vacuum solutions, with a
vanishing charge current density. \ The T-Spin vector vanishes identically,
but the T-Torsion vector does not. \ In fact, the T-Torsion vector is
constant. \ The solution corresponds to a circular state of polarization
with the constant magnetic and electric amplitudes rotating about the z
axis. \ The Poynting vector is not zero and is a constant, time independent,
vector. \ This wave solution is geometrically transverse (TEM), yet it
produces power as it is not topologically transverse (TTEM). \ If the
dispersion relation is not precisely satisfied, the current vector is
orthogonal to the T-Torsion vector and parallel to the vector potential. \
Both Poincare invariants vanish identically. The soliton like solution
should be compared to the wave guide solution of example 5 below, which is
also TEM, but does not radiate.

\subsection{Example 3. Complex Linear Polarization:}

Consider the Potentials

\begin{equation}
\ A=[\cos (kz-\omega t),i\cos (kz-\omega t),0,0]\,\ 
\end{equation}
and their induced fields:

\[
\mathbf{E}=\lbrack -\sin (kz-\omega t),-i\sin (kz-\omega t),0\rbrack \omega 
\]

\[
\mathbf{B}=\lbrack +i\sin (kz-\omega t),-\sin (kz-\omega t),0\rbrack k 
\]

\[
\mathbf{J}_{4}=\lbrack \cos (kz-\omega t),i\cos (kz-\omega t),0,0\rbrack
(k^{2}-\varepsilon \mu \omega ^{2})/\mu 
\]

\[
\mathbf{S}_{4}=\lbrack 0,0,0,0\rbrack . 
\]

\[
\mathbf{T}_{4}=\lbrack 0,0,0,0\rbrack . 
\]

\[
\mathbf{E\times H\,}=\lbrack 0,0,0\rbrack \, 
\]

\[
(\mathbf{B\circ H-D\circ E)}=0\mathbf{\,\,\,\,\,\,\,\,\,\,\,\,\,\,\,\,\,\,\,%
\,\,\,\,\,\,\,\,(E\circ B)}=0 
\]

This class of potentials generates a set of complex field intensities and
excitations, and a current density proportional to the vector potential. The
fields are said to be complex linearly polarized because the complex $%
\mathbf{B}$ field is a complex scalar multiple of the complex $\mathbf{E}$
field. \ If the dispersion relation $(k^{2}-\varepsilon \mu \omega ^{2})=0$
is satisfied, then the solutions are acceptable vacuum solutions, with a
vanishing charge current density. \ Note that both the T-Torsion vector and
the T-Spin vector vanish identically. \ The complex square of both the
electric and the magnetic field vectors vanish. \ Both Poincare invariants
vanish independent from the dispersion constraint. Although the fields are
propagating, there is no momentum flux and the Poynting vector is zero. \
The $\mathbf{E}$ and $\mathbf{B}$ fields are (complex) collinear. \ This
example is perhaps the simplest member of the class of Bateman-Whittaker
complex solutions described in Example 11, below.

\subsection{\protect\vspace{1pt}\qquad Example 4. Complex Circular
Polarization:}

Consider the Potentials

\begin{equation}
A=[\cos (kz-\omega t),i\sin (kz-\omega t),0,0]
\end{equation}
and their induced fields:

\[
\mathbf{E}=\lbrack -\sin (kz-\omega t),+i\cos (kz-\omega t),0\rbrack \omega 
\]

\[
\mathbf{B}=\lbrack -i\cos (kz-\omega t),-\sin (kz-\omega t),0\rbrack k 
\]

\[
\mathbf{J}_{4}=\lbrack \cos (kz-\omega t),i\sin (kz-\omega t),0,0\rbrack
(k^{2}-\varepsilon \mu \omega ^{2})/\mu 
\]

\[
\mathbf{S}_{4}=\lbrack 0,0,-k/\mu ,-\varepsilon \omega \rbrack \,2\cos
(kz-\omega t)\sin (kz-\omega t). 
\]

\[
\mathbf{T}_{4}=i\lbrack 0,0,-\omega ,-k\rbrack . 
\]

\[
\mathbf{E\times H\,}\,=\lbrack 0,0,-1\rbrack \,(\omega \,k/\mu )(2\cos
(kz-\omega t)^{2}-1) 
\]

\begin{eqnarray*}
(\mathbf{B\circ H-D\circ E)} &=&-2\{\cos (kz-\omega t)^{2}-\sin (kz-\omega
t)^{2}\}(k^{2}-\varepsilon \mu \omega ^{2})/\mu \mathbf{\,\,} \\
\mathbf{\,\,\,\,\,\,\,\,\,\,\,\,\,\,\,\,\,\,\,\,\,\,\,(E\circ B)} &=&0
\end{eqnarray*}

This class of potentials generates a set of complex field intensities and
excitations, and a current density proportional to the vector potential. \
If the dispersion relation $(k^{2}-\varepsilon \mu \omega ^{2})=0$ is
satisfied, then the solutions are acceptable vacuum solutions, with a
vanishing charge current density. Both the T-Torsion vector (imaginary) and
the T-Spin vector (real) do not vanish. \ The second Poincare invariant
vanishes identically, and the first Poincare invariant vanishes subject to
the dispersion constraint. \ The current vector, if non-zero due to
fluctuations in the dispersion relation, is orthogonal to both the T-Torsion
vector and the T-Spin vector.

\qquad Examples 1 through 4 above are geometrically transverse waves in the
engineering sense that the propagation direction of the phase (along the z
axis) is in the direction of the momentum flux, $\mathbf{D\times B.}$
However, the waves are not ''topologically transverse'' in that the sense
that the $\mathbf{D}$ and $\mathbf{B}$ fields are not necessarily transverse
to the components of the vector potential $\mathbf{A}.$

\subsection{Example 5. Waveguide TEM modes}

Consider the Potentials

\begin{equation}
\qquad A=[0,0,\phi (x,y),(\omega /k)\phi (x,y)]\cos (kz-\omega t)
\end{equation}
and their induced fields:

\[
\mathbf{E}=\lbrack -(\omega /k)\partial \phi /\partial x,-(\omega
/k)\partial \phi /\partial y,0\rbrack \cos (kz-\omega t) 
\]

\[
\mathbf{B}=\lbrack \partial \phi /\partial y,-\partial \phi /\partial
x,0\rbrack \cos (kz-\omega t) 
\]

\begin{eqnarray*}
\mathbf{J}_{4} &=&\lbrack \partial \phi /\partial x(\varepsilon \mu (\omega
/k)^{2}-1)\sin (kz-\omega t), \\
&&\partial \phi /\partial y(\varepsilon \mu (\omega /k)^{2}-1)\sin
(kz-\omega t), \\
&&\nabla ^{2}\phi \cos (kz-\omega t), \\
&&(\varepsilon \mu \omega /k)\nabla ^{2}\phi \cos (kz-\omega t)\rbrack /\mu
\end{eqnarray*}

\begin{eqnarray*}
\mathbf{S}_{4} &=&\lbrack \phi \partial \phi /\partial x\cos (kz-\omega
t)^{2}(1-\varepsilon \mu (\omega /k)^{2}), \\
&&\phi \partial \phi /\partial y\cos (kz-\omega t)^{2}(1-\varepsilon \mu
(\omega /k)^{2}), \\
&&0, \\
&&0\rbrack /\mu
\end{eqnarray*}

\[
\mathbf{T}_{4}=\lbrack 0,0,0,0\rbrack . 
\]

\begin{eqnarray*}
\mathbf{E\times H\,} &=&\lbrack \phi (\partial \phi /\partial x)k\cos
(kz-\omega t)\sin (kz-\omega t)(v_{g}-v_{p}), \\
&&\phi (\partial \phi /\partial y)k\cos (kz-\omega t)\sin (kz-\omega
t)(v_{g}-v_{p}), \\
&&(v_{g})\cos (kz-\omega t)^{2}(\nabla ^{2}\phi )\rbrack /\mu
\end{eqnarray*}

\[
\vspace{1pt}(\mathbf{B\circ H-D\circ E)}\neq 0\,\mathbf{\,\,\,\,\,\,\,\,\,\,%
\,\,\,\,\,\,\,\,\,\,\,\,\,(E\circ B)}=0 
\]

Note that the vector potential, $\mathbf{A,}$ is parallel to both the wave
vector, $\mathbf{k,}$ and the field momentum, $\mathbf{D}\times \mathbf{B.}$
\ The T-Torsion vector and the second Poincare invariant are identically
zero. \ The solution produces transverse current and T-Spin densities unless
a dispersion relation, $\varepsilon \mu (\omega /k)^{2}=1,$ is satisfied. \ $%
\,$Subject to the dispersion constraints, this classic solution has both a
zero T-Torsion vector and a zero T-Spin vector. \ Both $\mathbf{A\circ D}=0$
and $\mathbf{A\circ B}=0.$ \ The wave front is in the spatial direction of
the potential, by construction. \ The candidate solution subject to the
dispersion relation is both topologically transverse TTEM and geometrically
transverse, TEM .

\qquad However, even if the dispersion relations are satisfied, the
geometric TEM solution produces finite charge current densities, unless the
function $\phi (x,y)$ is a solution of the two dimensional Laplace equation, 
$\nabla ^{2}\phi =0.$ \ This further constraint implies that the TEM
solution produces no radiated power in the charge free state, for $\mathbf{%
E\times H}\,\Rightarrow 0$ \ as $\nabla ^{2}\phi \Rightarrow 0.$ \ In the
next example the constraint that the system be TTEM is relaxed, and radiated
power is achieved. in a TTM mode.

\subsection{\protect\vspace{1pt}\qquad Example 6. Waveguide TM modes}

\vspace{1pt}Consider the Potentials

\begin{equation}
\qquad A=[0,0,\phi (x,y)\cos (kz-\omega t),v_{g}\phi (x,y)\cos (kz-\omega t)
\end{equation}
\vspace{1pt}and their induced fields (note that example 6 differs from
example 5 in that a ''group'' velocity $v_{g}$ is used in the definition of
the potentials, instead of the phase velocity, $v_{p}=\omega /k)$:

\[
\mathbf{E}=\lbrack -v_{g}\partial \phi /\partial x,-v_{g}\partial \phi
/\partial y,\phi (x,y)\tan (kz-\omega t)(v_{g}k-\omega )\rbrack \cos
(kz-\omega t) 
\]

\[
\mathbf{B}=\lbrack \partial \phi (x,y)/\partial y\cos (kz-\omega
t),-\partial \phi (x,y)/\partial x\sin (kz-\omega t),0\rbrack 
\]

\begin{eqnarray*}
\mathbf{J}_{4} &=&\lbrack k\partial \phi /\partial x(\varepsilon \mu
v_{g}v_{p}-1)\sin ((kz-\omega t), \\
&&k\partial \phi /\partial y\sin ((kz-\omega t)(\varepsilon \mu
v_{g}v_{p}-1), \\
&&-(\nabla ^{2}\phi +\alpha \phi )\cos (kz-\omega t), \\
&&-v_{g}\varepsilon \mu (\nabla ^{2}\phi +\beta \phi )\cos (kz-\omega
t)\rbrack /\mu
\end{eqnarray*}

\[
\alpha =k^{2}\varepsilon \mu v_{p}(v_{p}-v_{g}),\,\,\,\,\,\,\,\,\,\,\,\beta
=k^{2}v_{g}(v_{p}/v_{g}-1) 
\]

\begin{eqnarray*}
\mathbf{S}_{4} &=&\lbrack -(\vspace{1pt}v_{g}/v_{p}-1)\phi \partial \phi
/\partial x\cos (kz-\omega t)^{2}, \\
&&-(\vspace{1pt}v_{g}/v_{p}-1)\phi \partial \phi /\partial x\cos (kz-\omega
t)^{2}, \\
&&-k(\vspace{1pt}v_{g}/v_{p}-1)\phi ^{2}\sin (kz-\omega t), \\
&&-\mu k(\vspace{1pt}v_{g}-v_{p})\phi ^{2}\sin (kz-\omega t)\rbrack /\mu
\end{eqnarray*}

\[
\mathbf{T}_{4}=\lbrack 0,0,0,0\rbrack . 
\]

\begin{eqnarray*}
\mathbf{E\times H} &=&\lbrack (\vspace{1pt}v_{p}/v_{g}-1)\phi \partial \phi
/\partial x\sin (kz-\omega t), \\
&&(\vspace{1pt}v_{p}/v_{g}-1)\phi \partial \phi /\partial y\sin (kz-\omega
t), \\
&&((\partial \phi /\partial x)^{2}+(\partial \phi /\partial y)^{2})\cos
(kz-\omega t)\rbrack (\,v_{g}/\mu )\cos (kz-\omega t)
\end{eqnarray*}

\begin{eqnarray*}
(\mathbf{B\circ H-D\circ E)} &=&-(\{\varepsilon \mu (\omega /k)^{2}-1\}/\mu
)\cos (kz-\omega t)^{2}\{(\nabla \phi )^{2}\mathbf{\,}+\phi (\nabla ^{2}\phi
)\} \\
\,\,\,\mathbf{\,\,\,\,\,\,\,\,\,\,\,\,\,\,\,\,\,\,\,\,(E\circ B)} &=&0
\end{eqnarray*}

\vspace{1pt}Note that in this solution, the fourth component of the Action
is scaled by the ''group velocity'', $v_{g},$ not the ''speed of light'', as
determined by the constitutive properties: $c=\sqrt{1/\xi \mu }.\,\,$This
class of potentials requires that the function $\phi (x,y)$ be a solution of
the two dimensional Helmholtz equation, $\nabla ^{2}\phi +\lambda ^{2}\phi
=0 $ . \ The phase velocity, $v_{p}=\omega /k,$ differs from the group
velocity, $v_{g}.$ \ Again, two constraint conditions (dispersion relations)
are required for the solution to be a vacuum solution without charge
currents. One of the constraint conditions demands that the product of the
group and the phase velocity, $v_{p}=\omega /k,$ to be equal to the square
of the speed of light as determined from the constitutive properties:

\begin{equation}
\vspace{1pt}v_{p}\cdot v_{g}=1/\varepsilon \mu =c^{2}.
\end{equation}

The second constraint required for the vacuum state $(\mathbf{J}=0,\rho =0)$
\ is determined by the Helmholtz parameter,\ $\lambda ,$and is satisfied when

\begin{equation}
\vspace{1pt}\lambda ^{2}=k^{2}(\vspace{1pt}v_{p}/v_{g}-1).
\end{equation}
Such TM modes are also TTM modes; \ the T-Torsion vector is identically
zero, but the T-Spin vector is not. \ Note that the solution becomes a TEM
mode solution when the phase velocity \ equals the group velocity, and the
function $\phi $ satisfies the Laplace equation, $\nabla ^{2}\phi =0.$ \
Further note that the $\mathbf{E}$ field has a longitudinal component when
the group velocity and the phase velocity are not the same. \ For the
transverse magnetic mode, $\mathbf{A\circ B}=0,\,$but $\mathbf{A\circ D}\neq
0$. The second Poincare invariant vanishes, $\mathbf{E\circ B}=0,$but for
this solution, the first Poincare invariant does not vanish. \ Not only is
the T-Spin vector not zero, but also its divergence is not zero. \ The
energy flow is in the direction of the wave vector, $\mathbf{k}$, but not in
the direction of the field momentum, $\mathbf{D\times B},\,$and the energy
propagates with the group velocity $v_{g}.$

\subsection{\qquad Example 7. An irreversible vacuum solution of type 1 for
which $\mathbf{E\circ B}\neq 0$}

Consider the potentials

\begin{equation}
\mathbf{A}=[+y,-x,ct]/\lambda ^{4}\,\,,\,\,\,\,\,\,\phi =cz/\lambda
^{4},\,\,\,\,\,where\,\,\lambda ^{2}=-c^{2}t^{2}+x^{2}+y^{2}+z^{2}.
\end{equation}
and their induced fields:

\[
\mathbf{E}=\lbrack
-2(cty-xz),+2(ctx+yz),-(c^{2}t^{2}+x^{2}+y^{2}-z^{2})\rbrack 2c/\lambda ^{6} 
\]

\[
\mathbf{B}=[-2(cty+xz),+2(ctx-yz),+(c^{2}t^{2}+x^{2}+y^{2}-z^{2})]2/\lambda
^{6}. 
\]

Subject to the dispersion relation, $\varepsilon \mu c^{2}=1.$ and the
Lorentz constitutive conditions, these time dependent wave functions satisfy
the homogeneous Maxwell equations without charge currents, and are therefore
acceptable vacuum solutions.

\begin{equation}
J_{+t}=dG=[0,0,0,0]
\end{equation}

The extensive algebra involved in these and other computations in this
article were checked with a Maple symbolic mathematics program [12]. It is
to be noted that when the substitution $t\Rightarrow -t$ is made in the
functional forms for the potentials, the modified potentials fail to satisfy
the vacuum Lorentz conditions for zero charge-currents. The algebraic
results for the charge current density are somewhat complicated, but the
bottom line is that

\begin{equation}
J_{-t}=dG\neq \lbrack 0,0,0,0].
\end{equation}

\textit{It appears that the valid vacuum solution presented above is not
time-reversal invariant.}

The T-Spin current density for this first non-transverse vacuum wave example
is evaluated as:

\begin{eqnarray}
Spin &:&\mathbf{S}_{4}=[x(3\lambda ^{2}-4y^{2}-4x^{2}),y(3\lambda
^{2}-4y^{2}-4x^{2}),z(\lambda ^{2}-4y^{2}-4x^{2}),  \nonumber \\
&&t(\lambda ^{2}-4y^{2}-4x^{2})](2/\mu )/\lambda ^{10},
\end{eqnarray}
and has zero divergence, subject to the condition $\varepsilon \mu c^{2}=1$.
\ Hence the first Poincare invariant is zero 
\begin{equation}
(\mathbf{B\circ H-D\circ E)}=0\mathbf{\,\,}
\end{equation}

The T-Torsion current may be evaluated as

\begin{equation}
T-Torsion:\mathbf{T}_{4}=-[x,y,z,t]2c/\lambda ^{8}.
\end{equation}
and has a non-zero divergence equal to the second Poincare invariant

\begin{equation}
Poincare\,\,2=-2\mathbf{E}\circ \mathbf{B=+}8c/\lambda ^{8}.
\end{equation}
The solution has magnetic helicity as $\mathbf{A}\circ \mathbf{B\neq }0$ and
is radiative in the sense that the Poynting vector, $\mathbf{E}\times 
\mathbf{H\neq }0\mathbf{.}$

Both the T-Spin current and the T-Torsion vector are non-zero, which implies
that this solution represents waves which are neither TTM nor TTE. \ They
are not transverse waves in any sense. \ However, the first Poincare
invariant vanishes, implying that the T-Spin integral is a deformation
invariant, and is conserved. \ The second Poincare invariant is not zero,
which implies that the T-Torsion-Helicity integral is not a topological
invariant. \ These solutions are not simple transverse waves for both $%
\mathbf{A\circ B}\neq 0\mathbf{,}$ and $\mathbf{A\circ D}\neq 0\mathbf{.}$
Note that the physical units of the second Poincare invariant are that of an
energy density multiplied by an impedance (ohms). \ As the second Poincare
invariant is not zero, it is impossible to find a compact without boundary
two surface that contains non-zero lines of magnetic field. \ That is, a
closed 2-torus of magnetic field lines does not exist.

\qquad However, as the first Poincare invariant is zero it is possible to
construct a deformation invariant in terms of the deRham period integral
over a closed 3 dimensional submanifold 
\begin{equation}
Spin=\vspace{1pt}\int_{z3}\{S_{x}dy\symbol{94}dz\symbol{94}dt-S_{y}dx\symbol{%
94}dz\symbol{94}dt+S_{z}dx\symbol{94}dy\symbol{94}dt-\sigma dx\symbol{94}dy%
\symbol{94}dz\}.
\end{equation}

\subsection{Example 8. \ An irreversible vacuum solution of type 2
complimentary to Example 7.}

Consider the potentials

\begin{equation}
\medskip \mathbf{A}=[+ct,-z,+y]/\lambda ^{4}\,\,,\,\,\,\,\,\,\phi
=cx/\lambda ^{4},\,\,\,\,\,where\,\,\lambda
^{2}=-c^{2}t^{2}+x^{2}+y^{2}+z^{2}
\end{equation}
and their induced fields:

\[
\mathbf{E}=\lbrack
+(-c^{2}t^{2}+x^{2}-y^{2}-z^{2}),+2(ctz+yx),-2(cty-zx)\rbrack 2c/\lambda
^{6} 
\]

\[
\mathbf{B}=[+(-c^{2}t^{2}+x^{2}-y^{2}-z^{2}),+2(-ctz+yx),+2(cty+zx)]2/%
\lambda ^{6}. 
\]
As in the previous example above, these fields satisfy the Maxwell-Faraday
equations, and the associated excitations satisfy the Maxwell-Ampere
equations without producing a charge current 4-vector. \ However, it follows
by direct computation that the second Poincare invariant, and the T-Torsion
4-vector are of opposite signs to the values computed for the previous
example:

\[
\mathbf{E}\circ \mathbf{B=+}4c/\lambda ^{8}\,\,\,\,\,\,and\,\,\,\ \ \ 
\mathbf{A}\circ \mathbf{B}=+2ct/\lambda ^{8}\,\,\,. 
\]

\subsection{\qquad Example 9. \ Superposition of the two complimentary
examples of type 1 and type 2.}

When the potentials of examples type 1 and type 2 above are combined by
addition or subtraction, the resulting wave is topologically transverse
magnetic, but not topological transverse electric. \ Not only does the
second Poincare invariant vanish under superposition, but so also does the
T-Torsion 4 vector. \ Conversely, the examples above show that there can
exist topologically transverse magnetic waves which can be decomposed into
two non-transverse waves. \ \thinspace\ A \ notable feature of the
superposed solutions is that the T-Spin 4 vector does not vanish, hence the
example superposition is a wave that is not topologically transverse
electric. \ However, for the examples above and their superposition, the
first Poincare invariant vanishes, which implies that the T-Spin remains a
conserved topological quantity for the superposition. The T-Spin current
density for the combined examples is given by the formula:

\begin{eqnarray}
\mathbf{S}_{4}
&=&[-2cx(y+ct)^{2},cy(y+ct)(x^{2}-y^{2}+z^{2}-2cty-c^{2}t^{2}),-2cz(y+ct)^{2},
\\
&&-(y+ct)(x^{2}+y^{2}+z^{2}+2cty+c^{2}t^{2})]4c/\lambda ^{10}  \nonumber
\end{eqnarray}
while the T-Torsion current is a zero vector

\[
\mathbf{T}_{4}=\lbrack 0,0,0,0\rbrack . 
\]

In addition, for the superposed example, the spatial components of the
Poynting vector are equal to the T-Spin current density vector multiplied by
\ $\gamma $, such that

\[
\mathbf{E\times H}=\gamma \,\,\mathbf{S},\,\,\,\,\,\ with\,\,\,\gamma
=-(x^{2}+y^{2}+z^{2}+2cty+c^{2}t^{2})/2c(y+ct)\lambda ^{2}. 
\]
These results seem to give classical credence to the Planck assumption that
vacuum state of \ Maxwell's electrodynamics supports quantized angular
momentum, (the conserved T-Spin integral) and that the energy flux must come
in multiples of the T-Spin quanta. \ In other words, these combined
irreversible solutions of examples type 1 and type 2 have the appearance of
the photon

\subsection{\qquad Example 10. \ Bateman-Whittaker solutions.}

\qquad In the modern language of differential forms, Bateman \lbrack
19\rbrack\ (and Whittaker) determined that if two \textit{complex} functions 
$\alpha (x,y,z,t)$ and $\beta (x,y,z,t)$ are used to define the 1-form of
Action,

\begin{equation}
A=\alpha d\beta -\beta d\alpha \Rightarrow \mathbf{A}=\alpha \nabla \beta
-\beta \nabla \alpha ,\,\,\,\,\,\,\phi =-(\alpha \partial \beta /\partial
t-\beta \partial \alpha /\partial t)
\end{equation}
then the derived 2-form \ $F=2d\alpha \symbol{94}d\beta $ generates the
complex field intensities,

\[
\mathbf{E}=(\partial \alpha /\partial t)\nabla \beta -(\partial \beta
/\partial t)\nabla \alpha \,\,\,\,and\,\,\,\,\,\,\mathbf{B}=\nabla \alpha
\times \nabla \beta , 
\]
which of course satisfy the Maxwell-Faraday equations. \ If in addition, the
functions $\alpha $ and $\beta $ satisfy the complex Bateman constraints:

\[
\nabla \alpha \times \nabla \beta =\pm (i/c)[(\partial \alpha /\partial
t)\nabla \beta -(\partial \beta /\partial t)\nabla \alpha ],\,\,\,\,\, 
\]
then the complex field excitations computed from the Lorentz vacuum
constitutive constraints will satisfy the Maxwell-Ampere equations for the
vacuum, without charge currents. \ It is apparent immediately that the
second Poincare invariant is identically zero for such solutions. \ It is
also apparent immediately that the T-Torsion vector is identically zero. \
What is not immediately apparent is that first Poincare invariant and the
T-Spin vector vanish identically as well. \ In fact, the constrained complex
solutions of the Bateman type are examples of topologically transverse
(TTEM) waves. \ \ The Bateman solutions do not radiate!

\qquad As an explicit example, consider

\[
\alpha =(x\pm iy)/(z-r),\,\,\,\,\,\beta =(r-ct),\,\ \ \ r=\sqrt{%
x^{2}+y^{2}+z^{2})}. 
\]
These functions satisfy the Bateman conditions (and, it should be mentioned,
the Eikonal equation subject to the dispersion relation $\varepsilon \mu
c^{2}=1).$ \ The $\mathbf{E}$ and the $\mathbf{B}$ fields are complex (and
complicated algebraically)

\begin{eqnarray*}
\vspace{1pt}\mathbf{B} &=&[yx+\sqrt{-1}(z^{2}+y^{2}-rz),-(z^{2}+x^{2}-rz)-%
\sqrt{-1}xy, \\
&&(r^{2}+z^{2}-2rz)/(r-z)\,)(y-\sqrt{-1}x)]2/(r(z-r)^{2})
\end{eqnarray*}

\begin{eqnarray*}
\vspace{1pt}\mathbf{E} &=&[-\sqrt{-1}yx+(y^{2}+z^{2}-rz),\sqrt{-1}%
(x^{2}+z^{2}-rz)-xy, \\
&&(z-r\,)(x+\sqrt{-1}y)]2c/(r(z-r)^{2})
\end{eqnarray*}

\[
\mathbf{S}_{4}=\lbrack 0,0,0,0\rbrack . 
\]

\[
\mathbf{T}_{4}=\lbrack 0,0,0,0\rbrack . 
\]

\[
\mathbf{E\times H}=\lbrack 0,0,0\rbrack ,\,\,\,\ \ \ \mathbf{D\times
B=\lbrack }0,0,0\rbrack ,\,\,\ \ \ \ \ \mathbf{E\circ E}=0,\mathbf{\,\,\ \ \
\,\,B\circ B}=0 
\]

\[
(\mathbf{B\circ H-D\circ E)}=0\mathbf{\,\,\,\,\,\,\,\,\,\,\,\,\,\,\,\,\,\,\,%
\,\,\,\,\,(E\circ B)}=0 
\]
The functions $\alpha $ and $\beta $ that satisfy the Bateman condition may
be used to construct an arbitrary function, $F(\alpha ,\beta ),$ and
remarkably enough, the arbitrary function $F(\alpha ,\beta )$ satisfies the
Eikonal equation,

\begin{equation}
(\nabla F)^{2}-\varepsilon \mu (\partial F/\partial t)^{2}=0.
\end{equation}

\vspace{1pt}From experience with Eikonal solutions and wave equations, it
might be thought that Eikonal solutions are sufficient. \ However, the
Bateman conditions are necessary, for both the candidate solutions

\begin{equation}
\alpha =(x\pm iy)/(z-ct),\,\,\,\,\,\beta =(r-ct),\,\ \ \ r=\sqrt{%
x^{2}+y^{2}+z^{2})}.
\end{equation}
satisfy the Eikonal equation, but not the Bateman conditions. \ They do not
generate TTEM modes in the vacuum. \ For arbitrary functions the algebra can
become quite complex. \ A Maple symbolic \ mathematics program for computing
the various terms is available (see references below)

\subsection{\qquad Example 11. \ A Plasma Accretion disc from HedgeHog B
field solutions.}

\qquad An interesting static solution that exhibits chiral symmetry breaking
can be obtained from the potentials

\begin{eqnarray}
\medskip \mathbf{A} &=&\Gamma (x,y,z,t)[-y,x,0]/(x^{2}+y^{2})\,\,,\,\ \  \\
with\,\ \,\,\,\,\Gamma &=&-z\,m/\sqrt{(x^{2}+y^{2}+\epsilon z^{2})} \\
and\,\,\,\,\,\,\,\phi &=&0.\,
\end{eqnarray}

These \ potentials induce the field intensities:

\begin{equation}
\mathbf{E}=[0,0,0]
\end{equation}

\begin{equation}
\mathbf{B}=m\,[x,y,z]/(x^{2}+y^{2}+\epsilon \,z^{2})^{3/2}.
\end{equation}
The $\mathbf{B}$ field is the famous Dirac Hedgehog field often associated
with ''magnetic monopoles''. \ \ However, the radial $\mathbf{B}$ field has
zero divergence everywhere except at the origin, which herein is interpreted
as a topological obstruction. \ The factor $\epsilon $ is to be interpreted
as an oblateness factor associated with rotation of a plasma, and is a
number between zero and 1. \ It is apparent that the helicity density and
the second Poincare invariant are zero:

\begin{equation}
\mathbf{E}\circ \mathbf{B}=0\,\,\,\,\,\,\,\,\,\,\,and\,\,\,\ \ \ \mathbf{A}%
\circ \mathbf{B}=0\,.
\end{equation}
In fact, the 3-form of T-Torsion vanishes identically (as $\phi =0),$

\begin{equation}
\mathbf{T}_{4}=[0,0,0,0].
\end{equation}

In this example, there is a non-zero value for the Amperian current density,
even though the potentials are static. \ The Current Density 3-form has
components,

\begin{equation}
\mathbf{J}_{4}=(3m/2\mu )\,\,(1-\epsilon
)\,z\,[-y,x,0,0]/(x^{2}+y^{2}+\epsilon \,z^{2})^{5/2}..
\end{equation}
which do not vanish if the system is ''oblate'' $(0<\epsilon <1).$ \ This
current density has a sense of \ ''circulation''\thinspace \thinspace about
the z axis, and is proportional to the vector potential reminiscent of a
London current, $\mathbf{J}=\lambda \mathbf{A}$. \ The ''order''\ parameter
is $(3/2\mu )\,\,(1-\epsilon )/(x^{2}+y^{2}+\epsilon \,z^{2})^{2}.$

\qquad The Lorentz force can be computed as:

\begin{equation}
\mathbf{J\times B=}(3m^{2}/4\mu )\,\,(1-\epsilon
)[xz^{2},yz^{2},-z]/(x^{2}+y^{2}+\epsilon \,z^{2})^{2}
\end{equation}
The formula demonstrates that the Lorentz force on the plasma, for the given
system of circulating currents, is directed radially away (centrifugally)
from the rotational axis, and yet is such that the plasma is attracted to
the $z=0,xy$ plane. \ The Lorentz force is divergent in the radial plane and
convergent in the direction of the z axis, towards the z=0 plane. \ This
electromagnetic field, therefor, would have the tendency to form an
accretion disk\ of the plasma in the presence of a central gravitational
field. \ 

\qquad Although the 3-form of T-Torsion vanishes identically, the 3-form of
T-Spin is not zero. The spatial components of the T-Spin are opposite to the
direction field of the Lorentz force (in the sense of a radiation reaction).
\ 

\begin{equation}
\mathbf{S}_{4}=(m^{2}/4\mu )[xz^{2},yz^{2},-z,0]/(x^{2}+y^{2}+\epsilon
\,z^{2})^{2}.
\end{equation}
The components of the T-Spin 3-form are in fact proportional to the
components of the virtual work 1-form. (See section 6) \ with the ratio $%
-3(1-\epsilon )$ depending on the oblateness factor.

It is also true that the divergence of the 3-form of T-Spin is not zero, for
the first Poincare invariant is

\begin{equation}
d(A\symbol{94}G)\Rightarrow P1=(m^{2}/4\mu )(x^{2}+y^{2}+4(1-\epsilon
)\,z^{2})/(x^{2}+y^{2}+\epsilon \,z^{2})^{3}
\end{equation}

\subsection{Example 11. \ Self dual solutions}

It is possible to construct a two-form $G$ \thinspace (without using the
Lorentz vacuum constitutive definitions) in terms of two arbitrary
functions, $\alpha $ and $\beta ,\,$from the dual relations:

\[
G=i(\ast d\alpha )\symbol{94}i(\ast d\beta )\Omega =i(\ast d\alpha )\symbol{%
94}i(\ast d\beta )dx\symbol{94}dy\symbol{94}dz\symbol{94}dt. 
\]
The functions $\alpha $ and $\beta $ used in the dual construction are not
required to be solutions of the Bateman condition. \ However, the resulting
''self-dual'' field excitations are \textbf{not} the same as those generated
by the Bateman method, unless the functions also satisfy the Bateman
conditions of complex collinearity. \ In the self dual formulas the *
operator is the Hodge * operator with respect to the Lorentz metric modified
by the impedance of free space. \ The resulting self-dual excitations
constructed from the two arbitrary functions indeed satisfy the
Maxwell-Ampere equations, in virtue of the Maxwell-Faraday equations and the
dispersion relation. \ The construction yields:

\[
\mathbf{H}=\sqrt{-1}/\mu c(\partial \alpha /\partial t)\nabla \beta
-(\partial \beta /\partial t)\nabla \alpha \,\,\,\,and\,\,\,\,\,\,\mathbf{D}%
=-\sqrt{-1}\varepsilon /c\nabla \alpha \times \nabla \beta . 
\]

\vspace{1pt}\vspace{1pt}The self-dual construction, however, implies a
chiral (non-Lorentz) constitutive relation of the type $\mathbf{D}=-[\gamma
]\circ \mathbf{B}$ and $\mathbf{H}=[\gamma ^{\dagger }]\circ \mathbf{E}$,
and will not be considered further in this article.

\section{SUMMARY}

\medskip

T-Torsion, $A\symbol{94}F,$ and T-Spin, $A\symbol{94}G$, have been
demonstrated to be useful theoretical concepts that give credence to the
physical reality of potentials in electromagnetic theory. \ The closed
integrals of these 3 forms and their divergences give precise meaning to the
concept of coherent structures in a plasma, and offer possible explanations
for certain astrophysical phenomena such as plasma jets from neutron stars,
and the formation of stable rings of material about rotating astrophysical
objects. \ The constructions can be used to define topological transverse
modes, similar to the geometric definitions of TM and TE modes.

\bigskip

\section{\protect\vspace{1pt}References}

\bigskip

\begin{enumerate}
\item  Bryant, R.L.,Chern, S.S., Gardner, R.B.,Goldschmidt, H.L., and
Griffiths, P. A. (1991), Exterior Differential Systems, Springer Verlag

\item  K. Kuratowski, Topology (Warsaw, 1948), Vol. I. \ \ \ 

Lipschutz, S. (1965) General Topology (Schaum, New York), 88.

\item  Cartan, E., (1958) Lecons sur les invariants integrauxs, Hermann,
Paris .

\item  Schouten, J. A. and Van der Kulk, W., (1949) Pfaff's Problem and its
Generalizations, Oxford Clarendon Press

\item  Kiehn, R. M. (1977) ''Periods on manfolds, quantization and gauge'',
J. of Math Phys \textbf{18}, no. 4, p. 614

\item  Kiehn, R. M., (1990) ''Topological Torsion, Pfaff Dimension and
Coherent Structures'', in: H. K. Moffatt and T. S. Tsinober eds, Topological
Fluid Mechanics, Cambridge University Press, 449-458 .

\item  deRham,G. (1960) Varietes Differentiables, Hermann, Paris

\item  TEM modes do not exist if the wave guide is simply connected.

\item  Hornig, G and Rastatter, L. (1998) ''The Magnetic Structure of B\ $%
\neq 0$ Reconnection'', Physica Scripta, Vol T74, p.34-39

\item  Sommerfeld, A., (1952) Electrodynamics, Academic Press, New York. \ 

Stratton, J.A., (1941) Electromagnetic Theory McGraw Hill N.Y.

Sommerfeld carefully distingushes between intensities and excitations on
thermodynamic grounds.

\item  Kiehn, R. M., Kiehn, G. P., and Roberds, R. B. (1991) ''Parity and
Time-reversal Symmetry Breaking, Singular Solutions'', Phys Rev A,\ \textbf{%
43}, p. 5665

\item  Kiehn, R. M. (1991) ''Are there three kinds of superconductivity''
Int. J. Mod. Phys B Vol. 5 p.1779-1790

Kiehn, R. M. and Pierce, J. F. (1969) ''An Intrinsic Transport Theorem''
Phys. Fluids \textbf{12}, p. 1971

\item  D. Van Dantzig, Proc. Cambridge Philos. Soc. \textbf{30}, 421 (1934).
Also see:

D. Van Dantzig, ''Electromagnetism Independent of metrical geometry'', Proc.
Kon. Ned. Akad. v. Wet. 37 (1934).

\item  Marsden, J.E. and Riatu, T. S.\ (1994) Introduction to Mechanics and
Symmetry, Springer-Verlag, p.122

\item  Kiehn, R. M. (1976) ''Retrodictive Determinism'', Int. J. of Eng.
Sci. \textbf{14}, p. 749

\item  K. Kuratowski, Topology (Warsaw, 1948), Vol. I.

\item  Lipschutz, S. (1965) General Topology (Schaum, New York), 88.

\item  Ibid 5

\item  Ibid 6

\item  Bateman, H. (1914, 1955)\ Electrical and Optical Wave Motion, Dover
p.12

\item  G. Hornig and K. Schindler, K. ''Magnetic topology and the problem of
its invariant definition'' Physics of Plasmas,\textbf{\ 3}, p.646 (1996).

\item  Kochin, N.E., Kibel, I.A., Roze, N. N., (1964) Theoretical
Hydrodynamics, Interscience, NY p.157

\item  Kiehn, R. M. (1975) ''Intrinsic hydrodynamics with applications to
space time fluids'', Int. J. Engng Sci \textbf{13}, p. 941-949
\end{enumerate}

\bigskip

\end{document}